\documentstyle[11pt]{article}
\begin{document}
\begin{titlepage}
\hfill{UQMATH-93-06}
\vskip.3in
\begin{center}
{\huge Quantum Affine Algebra and Universal $R$-Matrix with Spectral Parameter,
$II$}
\vskip.3in
{\Large Yao-Zhong Zhang} and {\Large Mark D. Gould}
\vskip.3in
{\large Department of Mathematics, University of Queensland, Brisbane,
Qld 4072, Australia}
\end{center}
\vskip.6in
\begin{center}
{\bf Abstract:}
\end{center}
This paper is an extended version of our previous short letter \cite{ZG2}
and is attempted to give a detailed account for the results presented in
that paper. Let $U_q({\cal G}^{(1)})$ be the quantized nontwisted affine
Lie algebra and $U_q({\cal G})$ be the corresponding quantum simple Lie
algebra. Using the previous obtained universal $R$-matrix for
$U_q(A_1^{(1)})$ and $U_q(A_2^{(1)})$, we determine the explicitly
spectral-dependent universal $R$-matrix for $U_q(A_1)$ and $U_q(A_2)$.
We apply these spectral-dependent universal $R$-matrix to some concrete
representations. We then reproduce the well-known results for the
fundamental representations and we are also able to derive for the first
time the extreamly explicit and compact formula of the spectral-dependent
$R$-matrix for the adjoint representation of $U_q(A_2)$, the
simplest nontrival case when the tensor product of the representations
is {\em not} multiplicity-free.

\end{titlepage}

\section{Introduction}
This paper is an extended version of our previous short letter \cite{ZG2}
where only the results have been announced and is partially attempted to
account for the details for the results presented in that letter.

Quantum deformation of universal enveloping algebras, or for short, quantum
algebra, is perhaps one of the most important discoveries in recent
years in mathematics and theoretical physics \cite{Drinfeld}\cite{Jimbo}.
The novelty in this theory is that it
has a quasitriangular Hopf algebra structure. Namely, there exists a canonical
element $R$, called the universal $R$-matrix, in the deformed algebra
satisfying the spectral-independent quantum Yang-Baxter equation (QYBE)
which plays a key role in CFT's \cite{Sierra} and knot theory
\cite{Reshetikhin}\cite{Witten}\cite{ZGB2}. Integrable models
\cite{Faddeev}\cite{Baxter}\cite{Wadati},
on the other hand, use spectral-dependent $R$-matrix which is the solution
to the spectral-dependent QYBE.

Since Jimbo and Jones' works \cite{Jimbo}\cite{Jones}, a central issue
has been finding spectral parameter dependent $R$-matrix using the
quantum group techiques (see, for example, \cite{ZGB}). The usual approach
seems to be Jones' "Yang-Baxterization" procedure. That is, given some
representation of braid group, one can in principal work out the
spectral-dependent solution to QYBE by Yang-Baxterizing the former.
This approach has been extensively applied to the case of so-called
"abelian Yang-Baxterization" where the tensor product of representations
is multiplicity-free. In fact, as far as we know, all the previous
research in literature has limited to this simple case. When the tensor
product of representations is {\em not} multiplicity-free, Jones
points out that "non-abelian Yang-Baxterization" plays a role. Therefore,
one may expect that one can not any more use the simple ansatz that
spectral-dependent $R$-matrix takes the form of spectral-dependent
scalar functions times spectral-independent projection operators, and
thus makes it very difficult to solve the Jimbo-type equations \cite{Jimbo}.

We will present a new way of obtaining the spectral-dependent $R$-matrix
for quantum simple Lie algebras. Our idea is, in some sense, to reverse
the above process. More precisely,
we start from the universal $R$-matrix of the quantum affine algebra
$U_q({\cal G}^{(1)})$ and then apply it to finite-dimensional loop
representations $V(z)$ of $U_q({\cal G}^{(1)})$ which is known to be isomorphic
to the ones, $V\otimes {\bf C}(z,z^{-1})$ of the corresponding quantum
simple Lie algebra $U_q({\cal G})$. In this way,
a spectral parameter appears automatically and we obtain the spectral
parameter dependent solution to QYBE for the latter. Our approach has
been partly initiated by Khoroshkin and Tolstoy's work \cite{KT} who consider
the simplest case of the fundametal representation of $U_q(A_1)$
and has classical analogue \cite{Bonora}. One of the advantages lying in our
method is that the multiplicity-free and {\em non}-multiplicity-free
cases can be treated in a unified fashion. As a matter of fact, we are
able to get a spectral-dependent universal $R$-matrix for $U_q(A_1)$
and $U_q(A_2)$. Applying to some concrete representations, we are
able to reproduce the well-known results for the fundamental
representations and to obtain for the first time the extreamly explicit and
compact formula for spectral-dependent $R$-matrix of $U_q(A_2)$ for the adjoint
representation, the simplest nontrival case when the tensor product of
representations is {\em not} multiplicity-free.

The present paper is set in the following fashion. In section 1 and 2,
we give some account for the fundamentals needed in this paper. In section
3, we give the universal $R$-matrix with the explicit spectral
dependence for $U_q(A_1)$ and $U_q(A_2)$. In section 4, we
apply the spectral-dependent universal $R$-matrix to some concrete
representations and reproduce some well-known results. We also
obtain an extreamly explicit and compact formula for spectral-dependent
$R$-matrix in the adjoint representation of $U_q(A_2)$. In
section 5 we present some remarks. Finally, some extra details are put in
Appendix A and B.

\section{Quantum Affine Lie Algebras}
\noindent
We start with the definition of the nontwisted quantum affine Lie algebra
$U_q({\cal G}^{(1)})$. Let $A^0=(a_{ij})_{1\leq i,j\leq r}$ be a
symmetrizable Cartan matrix. Let ${\cal G}$ stand for the finite-dimensional
simple Lie algebra associated with the symmetrical Cartan matrix
$A^0_{\rm sym}=(a^{\rm sym}_{ij})=(\alpha_i,\alpha_j),~i,j=1,2,...,r$,
where $r$ is the rank of ${\cal G}$.
Let $A=(a_{ij})_{0\leq i,j\leq r}$ be a symmetrizable,
generalized Cartan matrix in the sense of Kac \cite{Kac}.
Let ${\cal G}^{(1)}$ denote
the nontwisted affine Lie algebra associated with the corresponding symmetric
Cartan matrix $A_{\rm sym}=(a^{\rm sym}_{ij})=(\alpha_i,\alpha_j), ~i,j=0,1,
..., r$. Then the
quantum algebra $U_q({\cal G}^{(1)})$ is defined to be a Hopf algebra with
generators: $\{E_i,~F_i,~q^{h_i}~(i=0,1,...,r),~q^d\}$ and relations,
\begin{eqnarray}
&&q^h.q^{h'}=q^{h+h'}~~~~(h,~ h'=h_i~ (i=0,1,...,r),~d)\nonumber\\
&&q^hE_iq^{-h}=q^{(h,\alpha_i)} E_i\,,~~q^hF_iq^{-h}=
  q^{-(h,\alpha_i)}F_i\nonumber\\
&&[E_i, F_j]=\delta_{ij}\frac{q^{h_i}-q^{-h_i}}{q-q^{-1}}\nonumber\\
&&({\rm ad}_qE_i)^{1-a_{ij}}E_j=0\,,~~~({\rm ad}_{q^{-1}}F_i)^{1-a_{ij}}F_j=0
\,~~~~(i\neq j)\label{relations1}
\end{eqnarray}
where
\begin{equation}
({\rm ad}_qx_\alpha)x_\beta=[x_\alpha\,,\,x_\beta]_q=x_\alpha x_\beta -
  q^{(\alpha\,,\,\beta)}x_\beta x_\alpha\nonumber
\end{equation}

The algebra $U_q({\cal G}^{(1)})$ is a Hopf algebra with coproduct, counit and
antipode similar to the case of $U_q(\cal G)$: explicitly, the coproduct is
defined by
\begin{eqnarray}
&&\Delta(q^h)=q^h\otimes q^h\,,~~~h=h_i,~d\nonumber\\
&&\Delta(E_i)=q^{-h_i}\otimes E_i+E_i\otimes 1\nonumber\\
&&\Delta(F_i)=1\otimes F_i+F_i\otimes q^{h_i}\,,~~~i=0,1,...,r
\end{eqnarray}
Formulae for the counit and antipode may also be given, but are not required
below.

Let $\Delta'$ be the opposite coproduct: $\Delta'=T\,\Delta$,~$T(x\otimes
y)=y\otimes x\,,~\forall x,y\in U_q({\cal G}^{(1)})$. Then $\Delta$ and
$\Delta'$ is related by the universal  $R$-matrix $R$
in $U_q({\cal G}^{(1)})\otimes
U_q({\cal G}^{(1)})$ satisfying
\begin{eqnarray}
&&\Delta'(x)R=R\Delta(x)\,,~~~~~~x\in U_q({\cal G}^{(1)})\nonumber\\
&&(\Delta\otimes id )R=R^{13}R^{23}\,,~~~~(id\otimes\Delta)R=R^{13}R^{12}
\end{eqnarray}

We define an anti-involution $\theta$ on
$U_q({\cal G}^{(1)})$ by
\begin{equation}
\theta(q^h)=q^{-h}\,,~~\theta(E_i)=F_i\,,~~\theta(F_i)=E_i\,,~~\theta(q)=
  q^{-1}
\end{equation}
which extend uniquely to an algebra anti-involution on
all of $U_q({\cal G}^{(1)})$ so that
$\theta(ab)=\theta(b)\theta(a)\,,~~\forall a,b\in U_q({\cal G}^{(1)})$.
Throughout the paper, we use the notations:
\begin{eqnarray}
&&(n)_q=\frac{1-q^n}{1-q}\,,~~[n]_q=\frac{q^n-q^{-n}}{q-q^{-1}}\,,~~
  q_\alpha=q^{(\alpha,\alpha)}\nonumber\\
&&{\rm exp}_q(x)=\sum_{n\geq 0}\frac{x^n}{(n)_q!}\,,~~(n)_q!=
  (n)_q(n-1)_q\,...\,(1)_q
\end{eqnarray}

\section{Universal $R$-Matrix for $U_q(A_1^{(1)})$ and $U_q(A_2^{(1)})$}
This section is devoted to a brief review of the construction of the
universal $R$-matrix for $U_q(A_1^{(1)})$ and $U_q(A_2^{(1)})$ \cite{KT}
\cite{ZG}. We  start with  rank 2 case. Fix
a normal ordering in the positive root system $\Delta_+$ of $A_1^{(1)}$ :
\begin{equation}
\alpha,\,\alpha+\delta,\,...,\,\alpha+n\delta,\,...,\,\delta,\,2\delta,\,
...,\,m\delta,\,...\,,\,...\,,\,(\delta-\alpha)+l\delta,\,...\,,
\delta-\alpha\label{order1}
\end{equation}
where $\alpha$ and $\delta-\alpha$ are simple roots; $\delta$ is the
minimal positive imaginary root.
Construct Cartan-Weyl generators $E_\gamma\,,~F_\gamma=\theta(E_\gamma)
\,,~~\gamma\in \Delta_+$ of $U_q(A_1^{(1)})$ as follows:
We define
\begin{eqnarray}
&&\tilde{E_\delta}=[(\alpha,\alpha)]_q^{-1}[E_\alpha,\,E_{\delta-\alpha}]_q
  \nonumber\\
&&E_{\alpha+n\delta}=(-1)^n\left ({\rm ad}\tilde{E_\delta}\right )^nE_\alpha
  \nonumber\\
&&E_{(\delta-\alpha)+n\delta}=\left ({\rm ad}\tilde{E_\delta}\right )^n
  E_{\delta-\alpha}\,,...\nonumber\\
&&\tilde{E}_{n\delta}= [(\alpha,\alpha)]_q^{-1}[E_{\alpha+(n-1)\delta},
\,E_{\delta-\alpha}]_q  \label{cartan-weyl1}
\end{eqnarray}
where $[\tilde{E}_{n\delta},\,\tilde{E}_{m\delta}]=0$ for any $n,\,m >0$.
Then\\ (i) for
any $n>0$, there exists a unique element $E_{n\delta}$ \cite{KT} satisfying
$[E_{n\delta}\,,\,E_{m\delta}]=0$ for any $n,\,m>0$ and the relation
\begin{equation}
\tilde{E}_{n\delta}=\sum_{
\begin{array}{c}
k_1p_1+...+k_mp_m=n\\
0<k_1<...<k_m
\end{array}
}\frac{\left ( q^{(\alpha,\alpha)}-q^{-(\alpha,\alpha)}\right )^{\sum_ip_i-1}}
{p_1!\;...\;p_m!}(E_{k_1\delta})^{p_1}...(E_{k_m\delta})^{p_m}\label{ee1}
\end{equation}
(ii) the vectors $E_\gamma$ and $F_\gamma=
\theta(E_\gamma)$, $\gamma\in \Delta_+$ are the Cartan-Weyl generators for
$U_q(A^{(1)})$. One has \cite{KT}
\vskip.1in
\noindent {\bf Theorem:} The universal $R$-matrix for $U_q(A_1^{(1)})
$ may be written as
\begin{eqnarray}
R&=&\left (\prod_{n\geq 0}\;{\rm exp}_{q_\alpha}((q-q^{-1})(E_{\alpha+n\delta}
  \otimes  F_{\alpha+n\delta}))\right )\nonumber\\
& &\cdot{\rm exp}\left ( \sum_{n>0}n[n]_{q_\alpha}^{-1}
  (q_\alpha-q_\alpha^{-1})(E_{n\delta}\otimes F_{n\delta})\right )\nonumber\\
& &\cdot\left (\prod_{n\geq 0}\;{\rm exp}_{q_\alpha}((q-q^{-1})
  (E_{(\delta-\alpha)+n\delta}\otimes
  F_{(\delta-\alpha)+n\delta}))\right )\cdot
  q^{\frac{1}{2}h_\alpha\otimes h_\alpha+c\otimes d+d\otimes c}\label{sl2R}
\end{eqnarray}
where $c=h_\alpha+h_{\delta-\alpha}$. The order in the product
(\ref{sl2R}) concides
with the chosen normal order (\ref{order1}).

We now consider rank 3 case.
Let $A^0_{\rm sym}=(a^{\rm sym}_{ij})$,~~$i,j=1,2$ and $\Delta^0_+$
respectively be symmetrical Cartan matrix and positive root system of
rank 2 finite-dimensional simple Lie algebras $A_2$.
In what follows we use $A^0_{\rm sym}$ in the form
\begin{equation}
A^0_{\rm sym}=(a^{\rm sym}_{ij})=\left (
\begin{array}{cc}
(\alpha,\alpha) & (\alpha,\beta)\\
(\beta,\alpha) & (\beta,\beta)
\end{array}
\right )=\left (
\begin{array}{cc}
2 & -1\\
-1 & 2
\end{array} \right )
\end{equation}
The simple roots are $\alpha\,,\,\beta$ and $\delta-\psi$ with
$\psi=\alpha+\beta$ be the highest root of $A_2$.
\vskip.1in
\noindent {\bf Proposition:} For
$U_q(A^{(1)}_2)$, we fix the following order in $\Delta_+$ of $A_2^{(1)}$,
\begin{eqnarray}
&&\alpha,\,\alpha+\delta,\,...,\,\alpha+m_1\delta,\,...,\,\alpha+\beta,\,
\alpha+\beta+\delta,\,...,\,\alpha+\beta+m_2\delta,\,...,\,\beta,\,
\beta+\delta,\,...,\,\beta+m_3\delta,\,...,\,\delta,\,
2\delta,\,...,\,\nonumber\\
&&k\delta,\,...,\,...\,(\delta-\beta)+l_1\delta,\,...,\,\delta-\beta,\,...,\,
(\delta-\alpha)+l_2\delta,\,...,\,\delta-\alpha,\,...,\,(\delta-\alpha-\beta)
+l_3\delta,\,...,\,\delta-\alpha-\beta\nonumber\\
\label{ordering1}
\end{eqnarray}
where $m_i,k,l_i \geq 0\,,~~i=1,2,3$. We set
\begin{eqnarray}
&&E_{\alpha+\beta}=[E_\alpha\,,\,E_\beta]_q\,,~~~~~
  E_{\delta-\alpha}=[E_\beta\,,\,E_{\delta-\alpha-\beta}]_q\nonumber\\
&&E_{\delta-\beta}=[E_\alpha\,,\,E_{\delta-\alpha-\beta}]_q
\end{eqnarray}
and use the formula for $E_{\gamma+n\delta}$
and $E_{(\delta-\gamma)+n\delta}$, ~$\gamma\in \Delta_+^0$,
\begin{eqnarray}
&&\tilde{E}_\delta^{(i)}=[(\alpha_i,\alpha_i)]_q^{-1}[E_{\alpha_i},\,
  E_{\delta-\alpha_i}]_q\,,~~~\alpha_i=\alpha,\,\beta,\,\alpha+\beta\nonumber\\
&&E_{\alpha_i+n\delta}=(-1)^n\left ({\rm ad}\tilde{E}_\delta^{(i)}\right )^n
  E_{\alpha_i}\nonumber\\
&&E_{\delta-\alpha_i+n\delta}=\left ({\rm ad}\tilde{E}_\delta^{(i)}\right )^n
  E_{\delta-\alpha_i}\,,~~~~  ...\nonumber\\
&&\tilde{E}_{n\delta}^{(i)}= [(\alpha_i,\alpha_i)]_q^{-1}[E_{\alpha_i
  +(n-1)\delta},\,E_{\delta-\alpha_i}]_q  \label{cartan-weyl2}
\end{eqnarray}
where $[\tilde{E}^{(i)}_{n\delta},\,\tilde{E}^{(j)}_{m\delta}]=0$
for any $n,\,m >0$. One has the following statment similar to the case of
$U_q(A_1^{(1)})$ :\\
(i) there exists a unique element $E^{(i)}_{n\delta}\,,~n>0$ satisfying
$[E^{(i)}_{n\delta}\,,\,E^{(j)}_{m\delta}]=0$ for any $n,\,m>0$ and
the relation ($\alpha_i=\alpha,\,\beta$)
\begin{equation}
\tilde{E}^{(i)}_{n\delta}=\sum_{
\begin{array}{c}
k_1p_1+...+k_mp_m=n\\
0<k_1<...<k_m
\end{array}
}\frac{\left ( q^{(\alpha_i,\alpha_i)}-q^{-(\alpha_i,\alpha_i)}
\right )^{\sum_ip_i-1}}
{p_1!\;...\;p_m!}(E^{(i)}_{k_1\delta})^{p_1}...(E^{(i)}_{k_m\delta})^{p_m}
\label{ee2}
\end{equation}
(ii) the vectors $E_\gamma$ and $F_\gamma=
\theta(E_\gamma)$, $\gamma\in \Delta_+$ are the Cartan-Weyl generators for
$U_q(A_2^{(1)})$.\\
One can show \cite{KT}\cite{ZG} (see, in particular, \cite{ZG}) the following
\vskip.1in
\noindent{\bf Theorem:} For $U_q(A_2^{(1)})$, the universal $R$-matrix
takes the explicit form
\begin{eqnarray}
R&=&\prod_{n\geq 0}~{\rm exp}_{q_\alpha}
\left ((q-q^{-1})(E_{\alpha+n\delta}
\otimes F_{\alpha+n\delta})\right )\nonumber\\
& &\cdot\prod_{n\geq 0}~{\rm exp}_{q_{\alpha+\beta}}
\left ((q-q^{-1})(E_{\alpha+\beta+n\delta}
\otimes F_{\alpha+\beta+n\delta})\right )\nonumber\\
& &\cdot \prod_{n\geq 0}~{\rm exp}_{q_\beta}
\left ((q-q^{-1})(E_{\beta+n\delta}
\otimes F_{\beta+n\delta})\right )\nonumber\\
& &\cdot {\rm exp}\left (\sum_{n>0}\sum^2_{i,j=1}
C^n_{ij}(q)(q-q^{-1})(E^{(i)}_{n\delta}\otimes F^{(j)}_{n\delta})
\right )\nonumber\\
& &\cdot \prod_{n\geq 0}~{\rm exp}_{q_{\beta}}
\left ((q-q^{-1})(E_{(\delta-\beta)+n\delta}
\otimes F_{(\delta-\beta)+n\delta})\right )\nonumber\\
& &\cdot \prod_{n\geq 0}~{\rm exp}_{q_\alpha}
\left ((q-q^{-1})(E_{(\delta-\alpha)+n\delta}
\otimes F_{(\delta-\alpha)+n\delta})\right )\nonumber\\
& &\cdot\prod_{n\geq 0}~{\rm exp}_{q_{\alpha+\beta}}
\left ((q-q^{-1})(E_{(\delta-\alpha-\beta)+n\delta}
\otimes F_{(\delta-\alpha-\beta)+n\delta})\right )\nonumber\\
& &\cdot q^{\sum^2_{i,j=1}\,(a^{-1}_{\rm sym})^{ij}h_i\otimes h_j
+c\otimes d+d\otimes c}\label{aR}
\end{eqnarray}
where $c=h_0+h_\psi$, the order in the product of (\ref{aR})
is defined by (\ref{ordering1}) and the constants $C^n_{ij}(q)$ are given by
\begin{equation}
(C^n_{ij}(q))=(C^n_{ji}(q))=
 \frac{n}{[n]_q}\,\frac{[2]_q^2}{q^{2n}+1+q^{-2n}}\,\left (
\begin{array}{cc}
q^n+q^{-n} & (-1)^n\\
(-1)^n & q^n+q^{-n}
\end{array} \right )
\end{equation}

\section{Universal $R$-Matrix with Spectral Parameter}
In this section we come to our main concern. We will determine explicitly
spectral-dependent universal $R$-matrix for $U_q(A_1)$ and $U_q(A_2)$ by
using the universal $R$-matrix (\ref{sl2R}) and (\ref{aR}) for
the corresponding $U_q(A_1^{(1)})$ and
$U_q(A_2^{(1)})$, respectively.

We state the following
\vskip.1in
\noindent {\bf Lemma:} For any $z\in {\bf C}^\times$, there is a homomorphism
of algebras ${\rm ev}_z$: $U_q(A^{(1)}_1)\rightarrow U_q(A_1)$ given by
\begin{eqnarray}
&&{\rm ev}_z(E_\alpha)=E_\alpha\,,~~~{\rm ev}_z(F_\alpha)=F_\alpha\,,~~~
  {\rm ev}_z(h_\alpha)=h_\alpha\,,~~~{\rm ev}_z(c)=0\nonumber\\
&&{\rm ev}_z(E_\beta)=zF_\alpha\,,~~~{\rm ev}_z(F_\beta)=z^{-1}E_\alpha\,,~~~
  {\rm ev}_z(h_\beta)=-h_\alpha\label{lemma1}
\end{eqnarray}
\vskip.1in
\noindent {\bf Proof:} See the appendix A.
\vskip.1in
\noindent {\bf Proposition:} (Omitting "${\rm ev}_z$")
\begin{eqnarray}
&&E_{\alpha+n\delta}=(-1)^nz^nq^{-nh_\alpha}E_\alpha\nonumber\\
&&F_{\alpha+n\delta}=(-1)^nz^{-n}F_\alpha q^{nh_\alpha}\nonumber\\
&&E_{\beta+n\delta}=(-1)^nz^{n+1}F_\alpha q^{-nh_\alpha}\nonumber\\
&&F_{\beta+n\delta}=(-1)^nz^{-n-1}q^{nh_\alpha}E_\alpha\nonumber\\
&&\tilde{E}_{n\delta}=[2]_q^{-1}(-1)^{n-1}z^nq^{-(n-1)h_\alpha}\left (
  E_\alpha F_\alpha -q^{-2}F_\alpha E_\alpha\right )\nonumber\\
&&\tilde{F}_{n\delta}=[2]_q^{-1}(-1)^{n-1}z^{-n}q^{(n-1)h_\alpha}\left (
  F_\alpha E_\alpha -q^{2}E_\alpha F_\alpha\right )\label{proposition1}
\end{eqnarray}
\vskip.1in
\noindent {\bf Proof:} Straightforward calculations + induction in $n$, using
(\ref{lemma1}) and the defining relations (\ref{cartan-weyl1}).
\vskip.1in
We define the new primed quantities,
\begin{eqnarray}
&&\tilde{E}_{n\delta}\equiv z^n\,\tilde{E}'_{n\delta}\,,~~~~~~
  \tilde{F}_{n\delta}\equiv z^{-n}\,\tilde{F}'_{n\delta}\nonumber\\
&&E_{n\delta}\equiv z^n\,E'_{n\delta}\,,~~~~~~
  F_{n\delta}\equiv z^{-n}\,F'_{n\delta}
\end{eqnarray}
Then the tilded and primed quantities may be extracted from
(\ref{proposition1}) and
are obviously independent of the parameter $z$; moreover,
$E'_{n\delta}$ and $F'_{n\delta}$ are determined by
the following equalities of formal power series:
\begin{eqnarray}
&&(q_\alpha-q_\alpha^{-1})\sum_{k=1}^\infty \tilde{E}'_
{k\delta}u^k={\rm exp}\left ( (q_\alpha-q_\alpha^{-1})\sum_{l=1}^\infty
E'_{l\delta}u^l\right )-1\nonumber\\
&&-(q_\alpha-q_\alpha^{-1})\sum_{k=1}^\infty \tilde{F}'_
{k\delta}u^{-k}={\rm exp}\left ( -(q_\alpha-q_\alpha^{-1})\sum_{l=1}^\infty
F'_{l\delta}u^{-l}\right )-1\label{compare}
\end{eqnarray}
which are the variants of (\ref{ee1}). From the above considerations and
(\ref{sl2R}) we deduce
\vskip.1in
\noindent {\bf Theorem:}
The universal $R$-matrix of $U_q(A_1)$ with the explicit dependence
of spectral parameter, $R(x,y)\equiv ({\rm ev}_x\otimes {\rm ev}_y)R$,
can be written as the form,
\begin{eqnarray}
R(x,y)&=&\prod_{n\geq 0}\,{\rm exp}_{q_\alpha}\left ( (q-q^{-1})\left (
  \frac{x}{y}\right )^n\left (q^{-nh_\alpha}E_\alpha\otimes F_\alpha
  q^{nh_\alpha}\right )\right )\nonumber\\
& &\cdot{\rm exp}\left ( \sum_{n>0}n[n]_{q_\alpha}^{-1}
  (q_\alpha-q_\alpha^{-1})\left (\frac{x}{y}\right )^n
  (E'_{n\delta}\otimes F'_{n\delta})\right )\nonumber\\
& &\cdot\prod_{n\geq 0}\;{\rm exp}_{q_\alpha}\left ((q-q^{-1})\left (
  \frac{x}{y}\right )^{n+1}\left (F_\alpha q^{-nh_\alpha}\otimes q^{nh_\alpha}
  E_\alpha\right )\right )\cdot
  q^{\frac{1}{2}h_\alpha\otimes h_\alpha}\label{loop-sl2R}
\end{eqnarray}
\vskip.1in
\noindent We now consider the case of $U_q(A_2^{(1)})$. We state
\vskip.1in
\noindent {\bf Lemma:} For and $z\in {\bf C}^\times$, there is a homomorphism
of algebras ${\rm ev}_z$: $U_q(A_2^{(1)})\rightarrow U_q(A_2)$ given by
\begin{eqnarray}
&&{\rm ev}_z(E_\alpha)=E_\alpha\,,~~~{\rm ev}_z(F_\alpha)=F_\alpha\,,~~~
  {\rm ev}_z(h_\alpha)=h_\alpha\nonumber\\
&&{\rm ev}_z(E_\beta)=E_\beta\,,~~~{\rm ev}_z(F_\beta)=F_\beta\,,~~~
  {\rm ev}_z(h_\beta)=h_\beta\nonumber\\
&&{\rm ev}_z(E_{\delta-\alpha-\beta})=zF_{\alpha+\beta}q^{(h_\beta-h_\alpha)/3}
  \,,~~~{\rm ev}_z(F_{\delta-\alpha-\beta})=z^{-1}q^{(h_\alpha-h_\beta)/3}
  E_{\alpha+\beta}\nonumber\\
&&{\rm ev}_z(h_{\delta-\alpha-\beta})=-h_{\alpha+\beta}\,,~~~{\rm ev}_z(c)=0
  \label{lemma2}
\end{eqnarray}
\vskip.1in
\noindent {\bf Proof:} See the Appendix A.
\vskip.1in
\noindent {\bf Proposition:} (Omitting again "${\rm ev}_z$")
\begin{eqnarray}
&&E_{\alpha+n\delta}=(-1)^nz^nq^{-nh_\alpha}E_\alpha q^{-n(h_\alpha+2
  h_\beta)/3}\nonumber\\
&&F_{\alpha+n\delta}=(-1)^nz^{-n}q^{n(h_\alpha+2h_\beta)/3}F_\alpha
  q^{nh_\alpha}\nonumber\\
&&E_{\alpha+\beta+n\delta}=(-1)^nz^nq^{-nh_{\alpha+\beta}}E_{\alpha+\beta}
  q^{n(h_\beta-h_\alpha)/3}\nonumber\\
&&F_{\alpha+\beta+n\delta}=(-1)^nz^{-n}q^{n(h_\alpha-h_\beta)/3}F_{\alpha
  +\beta}q^{nh_{\alpha+\beta}}\nonumber\\
&&E_{\beta+n\delta}=(-1)^n[2]_q^{-n}z^nq^n\left \{\left ({\rm ad'}_{q^{-1}
  }{\cal E}\right )^nE_\beta\right \}q^{n(h_\beta-h_\alpha)/3}\nonumber\\
&&F_{\beta+n\delta}=[2]_q^{-n}z^{-n}q^{n(h_\alpha-h_\beta)/3}
  \left ({\rm ad'}_{q^{-1}}{\cal F}\right )^nF_\beta\nonumber\\
&&E_{(\delta-\beta)+n\delta}=[2]_q^{-n}z^{n+1}q^{-n}\left \{\left ({\rm ad'}_q
  {\cal E}\right )^n({\rm ad'}_{q^{-2}}E_\alpha)F_{\alpha+\beta}
  \right \}q^{(n+1)(h_\beta-h_\alpha)/3}\nonumber\\
&&F_{(\delta-\beta)+n\delta}=(-1)^n[2]_q^{-n}z^{-n-1}
  q^{(n+1)(h_\alpha-h_\beta)/3}
  \left ({\rm ad'}_q{\cal F}\right )^n({\rm ad'}_{q^2}E_{\alpha+\beta})F_\alpha
  \nonumber\\
&&E_{(\delta-\alpha)+n\delta}=(-1)^nz^{n+1}q^{-(n+1)(h_\alpha+2h_\beta)/3}
  F_\alpha q^{-nh_\alpha}\nonumber\\
&&F_{(\delta-\alpha)+n\delta}=(-1)^nz^{-n-1}q^{nh_\alpha}E_\alpha
  q^{(n+1)(h_\alpha+2h_\beta)/3}\nonumber\\
&&E_{(\delta-\alpha-\beta)+n\delta}=(-1)^nz^{n+1}q^{(n+1)(h_\beta-h_\alpha)/3}
  F_{\alpha+\beta} q^{-nh_{\alpha+\beta}}\nonumber\\
&&F_{(\delta-\alpha-\beta)+n\delta}=(-1)^nz^{-n-1}q^{nh_{\alpha+\beta}}
  E_{\alpha+\beta}q^{(n+1)(h_\alpha-h_\beta)/3}\nonumber\\
&&\tilde{E}^{(\alpha)}_{n\delta}=(-1)^{n-1}[2]^{-1}_qz^n(E_\alpha F_\alpha
  -q^{-2n}F_\alpha E_\alpha)\,q^{-(n-1)h_\alpha}q^{-n(h_\alpha+2h_\beta)/3}
  \nonumber\\
&&\tilde{F}^{(\alpha)}_{n\delta}=(-1)^{n-1}[2]^{-1}_qz^{-n}
  \,q^{(n-1)h_\alpha}q^{n(h_\alpha+2h_\beta)/3}(F_\alpha E_\alpha-q^{2n}
  E_\alpha F_\alpha)\nonumber\\
&&\tilde{E}^{(\beta)}_{n\delta}=(-1)^n[2]_q^{-n}z^nq^{n-2}\left \{
  \left ({\rm ad'}_{q^{-n+2}}{\cal F'}\right )\cdot\left ({\rm ad'}_{q^{-1}}
  {\cal E}\right )^{n-1}E_\beta\right \}q^{n(h_\beta-h_\alpha)/3}\nonumber\\
&&\tilde{F}^{(\beta)}_{n\delta}=[2]_q^{-n}z^{-n}q^{n-1}q^{n(h_\alpha-h_\beta)/3}
  \left ({\rm ad'}_{q^{-n+2}}{\cal E'}\right )\cdot
  \left ({\rm ad'}_{q^{-1}}{\cal F}\right )^{n-1}F_\beta\label{proposition2}
\end{eqnarray}
where
\begin{eqnarray}
&&({\rm ad'}_Q{\cal A})\cdot {\cal B}\equiv {\cal A}{\cal B}-Q{\cal B}{\cal A}
  \nonumber\\
&&{\cal E}=({\rm ad'}_{q^{-1}}E_\beta)({\rm ad'}_{q^{-2}}E_\alpha)
  F_{\alpha+\beta}\nonumber\\
&&{\cal F}=({\rm ad'}_q({\rm ad'}_{q^2}E_{\alpha+\beta})F_\alpha)F_\beta
  \nonumber\\
&&{\cal E'}=E_{\alpha+\beta}F_\alpha-q^2F_\alpha E_{\alpha+\beta}\nonumber\\
&&{\cal F'}=E_{\alpha}F_{\alpha+\beta}-q^{-2}F_{\alpha+\beta} E_{\alpha}
\end{eqnarray}
\vskip.1in
\noindent {\bf Proof:} Straightforward computations + induction in $n$,
by using (\ref{lemma2}) and the defining relations of generators,
eqs.(\ref{cartan-weyl2}) and (\ref{ee2}).
\vskip.1in
We define the primed quantities, motivated by the form of (\ref{proposition2}),
\begin{eqnarray}
&&\tilde{E}^{(\alpha)}_{n\delta}\equiv z^n\tilde{E}^{'(\alpha)}_{n\delta}\,,
  ~~~~\tilde{F}^{(\alpha)}_{n\delta}\equiv z^{-n}
  \tilde{F}^{'(\alpha)}_{n\delta}\nonumber\\
&&\tilde{E}^{(\beta)}_{n\delta}\equiv z^n\tilde{E}^{'(\beta)}_{n\delta}\,,
  ~~~~\tilde{F}^{(\beta)}_{n\delta}\equiv z^{-n}
  \tilde{F}^{'(\beta)}_{n\delta}\nonumber\\
&&E^{(\alpha)}_{n\delta}\equiv z^nE^{'(\alpha)}_{n\delta}\,,
  ~~~~F^{(\alpha)}_{n\delta}\equiv z^{-n}
  F^{'(\alpha)}_{n\delta}\nonumber\\
&&E^{(\beta)}_{n\delta}\equiv z^nE^{'(\beta)}_{n\delta}\,,
  ~~~~F^{(\beta)}_{n\delta}\equiv z^{-n}
  F^{'(\beta)}_{n\delta}\nonumber\\
&&E_{\beta+n\delta}\equiv z^n E'_{\beta+n\delta}\,,~~~~
  F_{\beta+n\delta}\equiv z^{-n}F'_{\beta+n\delta}\nonumber\\
&&E_{(\delta-\beta)+n\delta}\equiv z^{n+1}E'_{(\delta-\beta)+n\delta}
  \nonumber\\
&&F_{(\delta-\beta)+n\delta}\equiv z^{-n-1}F'_{(\delta-\beta)+n\delta}
  \label{primed2}
\end{eqnarray}
Then the primed quantities do not depend on the parameter $z$
and $\tilde{E}^{'(i)}_{n\delta}$\,,~$\tilde{F}^{'(i)}_{n\delta}$\,,~
$E'_{\beta+n\delta}$\,,~$F'_{\beta+n\delta}$\,,~$E'_{(\delta-\beta)+n\delta}$
\,,~$F'_{(\delta-\beta)+n\delta}$ can be easily read off by comparing
(\ref{primed2} with (\ref{proposition2}); moreover, similar to the
$U_q(A_1^{(1)})$ case, $E^{'(i)}_{n\delta}\,,~~F^{'(i)}_{n\delta}$ are
determined by the following
equalities of formal series: ($\alpha_i=\alpha,\,\beta$)
\begin{eqnarray}
&&(q_{\alpha_i}-q_{\alpha_i}^{-1})\sum_{k=1}^\infty\tilde{E}^{'(i)}_
  {k\delta}u^k={\rm exp}\left ((q_{\alpha_i}-q_{\alpha_i}^{-1})
  \sum_{l=1}^\infty E^{'(i)}_{l\delta}u^l\right )-1\nonumber\\
&&-(q_{\alpha_i}-q_{\alpha_i}^{-1})\sum_{k=1}^\infty\tilde{F}^{'(i)}_
  {k\delta}u^{-k}={\rm exp}\left (-(q_{\alpha_i}-q_{\alpha_i}^{-1})
  \sum_{l=1}^\infty F^{'(i)}_{l\delta}u^{-l}\right )-1\label{primed}
\end{eqnarray}
which are the variants of (\ref{ee2}). The above considerations and
(\ref{aR}) then give rise to
\vskip.1in
\noindent {\bf Theorem:} The universal $R$-matrix of $U_q(A_2)$
with the explicit dependence of spectral parameter, $R(x,y)\equiv
({\rm ev}_x\otimes {\rm ev}_y)R$, takes the form,
\begin{eqnarray}
R(x,y)&=&\prod_{n\geq 0}~{\rm exp}_{q_\alpha}
\left ((q-q^{-1})\left (\frac{x}{y}\right )^n\left (q^{-nh_\alpha}E_\alpha
q^{-n(h_\alpha+2h_\beta)/3}\otimes q^{n(h_\alpha+2\beta)/3}F_\alpha
q^{nh_\alpha}\right )\right )\nonumber\\
& &\cdot\prod_{n\geq 0}~{\rm exp}_{q_{\alpha+\beta}}
\left ((q-q^{-1})\left (\frac{x}{y}\right )^n\left (q^{-nh_{\alpha+\beta}}
E_{\alpha+\beta}q^{n(h_\beta-h_\alpha)/3}
\otimes q^{n(h_\alpha-\beta)/3}F_{\alpha+\beta}q^{nh_{\alpha+\beta}}
\right )\right )\nonumber\\
& &\cdot \prod_{n\geq 0}~{\rm exp}_{q_\beta}
\left ((q-q^{-1})\left (\frac{x}{y}\right )^n\left (E'_{\beta+n\delta}
\otimes F'_{\beta+n\delta}\right )\right )\nonumber\\
& &\cdot {\rm exp}\left (\sum_{n>0}\sum^2_{i,j=1}
C^n_{ij}(q)(q-q^{-1})(E^{'(i)}_{n\delta}\otimes F^{'(j)}_{n\delta})
\right )\nonumber\\
& &\cdot \prod_{n\geq 0}~{\rm exp}_{q_{\beta}}
\left ((q-q^{-1})\left (\frac{x}{y}\right )^{n+1}\left (E'_{(\delta-\beta)
+n\delta}\otimes F'_{(\delta-\beta)+n\delta}\right )\right )\nonumber\\
& &\cdot \prod_{n\geq 0}~{\rm exp}_{q_\alpha}
\left ((q-q^{-1})\left (\frac{x}{y}\right )^{n+1}\left (
q^{-(n+1)(h_\alpha+2h_\beta)/3}F_\alpha q^{-nh_\alpha}
\otimes q^{nh_\alpha}E_\alpha q^{(n+1)(h_\alpha+2h_\beta)/3}
\right )\right )\nonumber\\
& &\cdot\prod_{n\geq 0}~{\rm exp}_{q_{\alpha+\beta}}
\left ((q-q^{-1})\left (\frac{x}{y}\right )^{n+1}\left (q^{(n+1)(h_\beta-
h_\alpha)/3}F_{\alpha+\beta}q^{-nh_{\alpha+\beta}}\right .\right .\nonumber\\
& &\left .\left .~~~~~
\otimes q^{nh_{\alpha+\beta}}E_{\alpha+\beta}q^{(n+1)(h_\alpha-h_\beta)/3}
\right )\right )
\cdot q^{\sum^2_{i,j=1}\,(a^{-1}_{\rm sym})^{ij}h_i\otimes h_j}\label{loop-R}
\end{eqnarray}

\section{Applications}
To illustrate the general theory developed in the previous section, we
present a detailed study of the spectral-dependent $R$-matrix for
some concrete and interesting representations.

First consider the $U_q(A_1^{(1)})$ case.
Let $V_l\,,~l\in {\bf Z_+}$ denote the $(l+1)$-dimensional module of $U_q(A_1)$
(spin $l/2$ representation) with basis $\{v^{(l)}_m\,|\,0\leq m\leq l\}$.
We have
\vskip.1in
\noindent {\bf Proposition:} For spin $l/2$ representation of $U_q(A_1)$,
we have
\begin{eqnarray}
h_\alpha v^{(l)}_m&=&(l-2m)v^{(l)}_m\nonumber\\
E_\alpha v^{(l)}_m&=&[l-m+1]_qv^{(l)}_{m-1}\nonumber\\
F_\alpha v^{(l)}_m&=&[m+1]_qv^{(l)}_{m+1}\nonumber\\
E'_{n\delta}v^{(l)}_m&=&[2]_q^{-1}\frac{(-1)^{n-1}}{n}q^{nm}\left (
  [n(l-m)]_q-q^{-n(l+2)}[nm]_q\right ) v^{(l)}_m\nonumber\\
F'_{n\delta}v^{(l)}_m&=&[2]_q^{-1}\frac{(-1)^{n-1}}{n}q^{-nm}\left (
  [n(l-m)]_q-q^{n(l+2)}[nm]_q\right ) v^{(l)}_m\label{loop-reps}
\end{eqnarray}
where it is understood that $v^{(l)}_m$ is identically zero if $m>l$ or
$m<0$.
\vskip.1in
\noindent {\bf Proof:} Straightforward computations + induction in $n$.
\vskip.1in
\noindent (i) for spin $1/2$ representation, we have from
(\ref{loop-reps})
\begin{eqnarray}
&&h_\alpha=\left (
\begin{array}{cc}
1 & 0\\
0 & -1
\end{array}
\right )\,,~~~E_\alpha=\left (
\begin{array}{cc}
0 & 1\\
0 & 0
\end{array}
\right )\,,~~~F_\alpha=\left (
\begin{array}{cc}
0 & 0\\
1 & 0
\end{array}
\right )\nonumber\\
&&E'_{n\delta}=[2]^{-1}_q\frac{(-1)^{n-1}}{n} [n]_q\left (
\begin{array}{cc}
1 & 0\\
0 & -q^{-2n}
\end{array}
\right )\,,~~~F'_{n\delta}=[2]^{-1}_q\frac{(-1)^{n-1}}{n} [n]_q\left (
\begin{array}{cc}
1 & 0\\
0 & -q^{2n}
\end{array}
\right )\label{spin 1/2}
\end{eqnarray}

We apply (\ref{loop-sl2R}) to $V_{1/2}\otimes V_{1/2}$, where
$V_{1/2}$ denotes the spin-$1/2$ representation of $U_q(A_1)$.
Using (\ref{spin 1/2}), it follows from (\ref{loop-sl2R}) that
\begin{equation}
R_{1/2,1/2}(x,y)=f_q(x,y)\cdot \left (
\begin{array}{cccc}
1 & {} & {} & {}\\
{} & \frac{q^{-1}(y-x)}{y-q^{-2}x} & \frac{y(1-q^{-2})}{y-q^{-2}x} & {}\\
{} & \frac{x(1-q^{-2})}{y-q^{-2}x} & \frac{q^{-1}(y-x)}{y-q^{-2}x} & {}\\
{} & {} & {} & 1
\end{array}
\right )\label{1/2,1/2}
\end{equation}
where
\begin{equation}
f_q(x,y)=q^{1/2}\cdot {\rm exp}\left (\sum_{n>0}\frac{q^n-q^{-n}}{q^n+q^{-n}}
\frac{(x/y)^n}{n}\right )
\end{equation}
and  use has been made of the notation:
\begin{equation}
(A\otimes B)=\left (
\begin{array}{cccc}
A_{11}B & A_{12}B & \cdots & A_{1N}B\\
. & . & . & .\\
. & . & . & .\\
A_{M1}B & A_{M2}B & \cdots & A_{MN}B
\end{array}
\right )
\end{equation}
We thus reproduce the well-known result \cite{Jimbo}, up to a scalar factor
$f_q(x,y)$.
In \cite{KT}, KT obtained (\ref{1/2,1/2}) directly from (\ref{sl2R}).\\
(ii) for spin $1$ representation, (\ref{loop-reps}) give rise to
\begin{eqnarray}
&&h_\alpha=\left (
\begin{array}{ccc}
2 & 0 & 0\\
0 & 0 & 0\\
0 & 0 & -2
\end{array}
\right )\,,~~~E_\alpha=\left (
\begin{array}{ccc}
0 & [2]_q & 0\\
0 & 0 & 1\\
0 & 0 & 0
\end{array}
\right )\,,~~~F_\alpha=\left (
\begin{array}{ccc}
0 & 0 & 0\\
1 & 0 & 0\\
0 & [2]_q & 0
\end{array}
\right )\nonumber\\
&&F'_{n\delta}=[2]_q^{-1}\frac{(-1)^{n-1}}{n}[n]_q\left (
\begin{array}{ccc}
q^n+q^{-n} & 0 & 0\\
0 & -q^n(q^{2n}-q^{-2n}) & 0\\
0 & 0 & -q^{2n}(q^n+q^{-n})
\end{array}
\right )\label{spin 1}
\end{eqnarray}

We now apply (\ref{loop-sl2R}) to $V_{1/2}\otimes V_1$ with
$V_1$ being the spin-$1$ representation of $U_q(A_1)$.
Using (\ref{spin 1}), we obtain from (\ref{loop-sl2R}),
\begin{eqnarray}
R_{1/2,1}(x,y)&=&\frac{q^{2}(y-q^{-1}x)}{y-qx}\cdot\left (e_{11}+e_{66}
 +\frac{q^{-2}(y-qx)}{y-q^{-3}x}(e_{33}+e_{44})+\right .\nonumber\\
& &\left .+\frac{yq^{-1}(1-q^{-2})}{y-q^{-3}x}e_{24}+\frac{xq^{-1}(1-q^{-2})}
 {y-q^{-3}x}e_{53}\right )\label{1/2 1}
\end{eqnarray}
where $e_{ij}$ is the matrix satisfying $(e_{ij})_{kl}=\delta_{ik}
\delta_{jl}$ and $e_{ij}e_{kl}=\delta_{jk}e_{il}$.

We now turn to the $U_q(A_2^{(1)})$ case. We state
\vskip.1in
\noindent {\bf Proposition:} The explicit form of generators on the fundamental
representation of $U_q(A_2)$ is given by
\begin{eqnarray}
&&h_\alpha={\rm diag}(1,-1,0)\,,~~~~~~h_\beta={\rm diag}(0,1,-1)\nonumber\\
&&E_\alpha=\left (
\begin{array}{ccc}
0 & 1 & 0\\
0 & 0 & 0\\
0 & 0 & 0
\end{array}
\right )\,,~~~~~~F_\alpha=\left (
\begin{array}{ccc}
0 & 0 & 0\\
1 & 0 & 0\\
0 & 0 & 0
\end{array}
\right )\nonumber\\
&&E_{\alpha+\beta}=\left (
\begin{array}{ccc}
0 & 0 & 1\\
0 & 0 & 0\\
0 & 0 & 0
\end{array}
\right )\,,~~~~F_{\alpha+\beta}=\left (
\begin{array}{ccc}
0 & 0 & 0\\
0 & 0 & 0\\
1 & 0 & 0
\end{array}
\right )\nonumber\\
&&E'_{\beta+n\delta}=q^{-2n-n/3}\left (
\begin{array}{ccc}
0 & 0 & 0\\
0 & 0 & 1\\
0 & 0 & 0
\end{array}
\right )\,,~~F'_{\beta+n\delta}=q^{2n+n/3}\left (
\begin{array}{ccc}
0 & 0 & 0\\
0 & 0 & 0\\
0 & 1 & 0
\end{array}
\right )\nonumber\\
&&E'_{(\delta-\beta)+n\delta}=(-1)^nq^{-2n-n/3-1}\left (
\begin{array}{ccc}
0 & 0 & 0\\
0 & 0 & 0\\
0 & 1 & 0
\end{array}
\right )\nonumber\\
&&F'_{(\delta-\beta)+n\delta}=(-1)^nq^{2n+n/3+1}\left (
\begin{array}{ccc}
0 & 0 & 0\\
0 & 0 & 1\\
0 & 0 & 0
\end{array}
\right )\nonumber\\
&&E^{'(\alpha)}_{n\delta}=[2]_q^{-1}(-1)^{n-1}\frac{[n]_q}{n}q^{-n/3}
  \,{\rm diag}\left (1,-q^{-2n},0\right )\nonumber\\
&&F^{'(\alpha)}_{n\delta}=[2]_q^{-1}(-1)^{n-1}\frac{[n]_q}{n}q^{n/3}
  \,{\rm diag}\left (1,-q^{2n},0\right )\nonumber\\
&&E^{'(\beta)}_{n\delta}=[2]_q^{-1}\frac{[n]_q}{n}q^{-n-n/3}
  \,{\rm diag}\left (0,-1,q^{-2n}\right )\nonumber\\
&&F^{'(\beta)}_{n\delta}=[2]_q^{-1}\frac{[n]_q}{n}q^{n+n/3}
  \,{\rm diag}\left (0,-1,q^{2n}\right )\label{proposition3}
\end{eqnarray}
\vskip.1in
We apply (\ref{loop-R}) to $V_{(3)}\otimes V_{(3)}$, where $V_{(3)}$
stands for the fundamental representation of $U_q(A_2)$. Using
(\ref{proposition3}) we get from (\ref{loop-R}),
It follows from (\ref{loop-R}) that
\begin{eqnarray}
R_{(3),(3)}(x,y)&=&f_q(x,y)\cdot \left (e_{11}+e_{99}
  +\frac{q^{-1}(y-x)}{y-q^{-2}x}(e_{22}+e_{33}+e_{44}+e_{66}+e_{77}+e_{88})
  +\right .\nonumber\\
& &\left .+\frac{y(1-q^{-2})}{y-q^{-2}x}(e_{24}+e_{37}+e_{68})
  +\frac{x(1-q^{-2})}{y-q^{-2}x}(e_{42}+e_{73}+e_{86})\right )
\end{eqnarray}
where
\begin{equation}
f_q(x,y)=q^{2/3}\cdot {\rm exp}\left (\sum_{n>0}\frac{q^{2n}-q^{-2n}}{q^{2n}+1
 +q^{-2n}}\,\frac{(x/y)^n}{n}\right )
\end{equation}
We thus reproduce the well-known result \cite{Jimbo}, up to a scalar factor
$f_q(x,y)$.

We now consider a very interesting case: to extract the spectral depedendent
$R$-matrix in the adjoint representation of $U_q(A_2)$.
As one knows, this is simplest nontrivial
example where the tensor product is not multiplicity-free. To this effect,
we introduce the so-called Gelfand-Tsetlin basis vector $|(m)>$ given by
\begin{equation}
|(m)>=\left |\left (
\begin{array}{c}
m_{13}~~~m_{23}~~~m_{33}\\
m_{12}~~~m_{22}\\
m_{11}
\end{array}
\right )\right >
\end{equation}
It can be shown that the action of generators on the GT basis vectors reads
\begin{eqnarray*}
&&h_\alpha|(m)>=(2m_{11}-m_{12}-m_{22})|(m)>\nonumber\\
&&h_\beta|(m)>=(2m_{12}+2m_{22}-m_{11}-m_{13}-m_{23}-m_{33})|(m)>\nonumber\\
\end{eqnarray*}
\begin{eqnarray}
&&F_\alpha\left |\left (
\begin{array}{c}
m_{13}~~~m_{23}~~~m_{33}\\
m_{12}~~~m_{22}\\
m_{11}
\end{array}
\right )\right >=\left \{[m_{11}-m_{22}]_q[m_{12}-m_{11}+1]_q\right \}^{1/2}
\left |\left (
\begin{array}{c}
m_{13}~~~m_{23}~~~m_{33}\\
m_{12}~~~m_{22}\\
m_{11}-1
\end{array}
\right )\right >\nonumber\\
&&F_\beta\left |\left (
\begin{array}{c}
m_{13}~~~m_{23}~~~m_{33}\\
m_{12}~~~m_{22}\\
m_{11}
\end{array}
\right )\right >=\left \{\frac{[m_{12}-m_{11}]_q[m_{13}-m_{12}+1]_q
  [m_{23}-m_{12}]_q[m_{33}-m_{12}-1]_q}{[m_{12}-m_{22}+1]_q
  [m_{12}-m_{22}]_q}\right \}^{1/2}\nonumber\\
&&~~~~~~~~~~~~~~~~~~~~\times \left |\left (
\begin{array}{c}
m_{13}~~~m_{23}~~~m_{33}\\
m_{12}-1~~~m_{22}\\
m_{11}
\end{array}
\right )\right >+\nonumber\\
&&~~~~~~~~~~~~~~~~~~~~~+\left \{\frac{[m_{22}-m_{11}-1]_q[m_{13}-m_{22}+2]_q
  [m_{23}-m_{22}+1]_q
  [m_{33}-m_{22}]_q}{[m_{12}-m_{22}+2]_q[m_{12}-m_{22}+1]_q}\right \}
  ^{1/2}\nonumber\\
&&~~~~~~~~~~~~~~~~~~~~~\times \left |\left (
\begin{array}{c}
m_{13}~~~m_{23}~~~m_{33}\\
m_{12}~~~m_{22}-1\\
m_{11}
\end{array}
\right )\right >
\end{eqnarray}
The matrix elements of $E_\alpha$ and $E_\beta$ are given by the transpose
of the ones of $F_\alpha$ and $F_\beta$, respectively. Now for the adjoint
representation, we have the following  $8$ state vectors:
\begin{eqnarray}
&&\phi_1=\left |\left (
\begin{array}{c}
1~~~0~~~-1\\
1~~~0\\
1
\end{array}
\right )\right >\,,~~~\phi_2=\left |\left (
\begin{array}{c}
1~~~0~~~-1\\
1~~~0\\
0
\end{array}
\right )\right >\,,~~~\phi_3=\left |\left (
\begin{array}{c}
1~~~0~~~-1\\
1~~~-1\\
1
\end{array}
\right )\right >\nonumber\\
&&\phi_4=\left |\left (
\begin{array}{c}
1~~~0~~~-1\\
1~~~-1\\
0
\end{array}
\right )\right >\,,~~~\phi_5=\left |\left (
\begin{array}{c}
1~~~0~~~-1\\
0~~~0\\
0
\end{array}
\right )\right >\,,~~~\phi_6=\left |\left (
\begin{array}{c}
1~~~0~~~-1\\
1~~~-1\\
-1
\end{array}
\right )\right >\nonumber\\
&&\phi_7=\left |\left (
\begin{array}{c}
1~~~0~~~-1\\
0~~~-1\\
0
\end{array}
\right )\right >\,,~~~\phi_8=\left |\left (
\begin{array}{c}
1~~~0~~~-1\\
0~~~-1\\
-1
\end{array}
\right )\right >
\end{eqnarray}
Therefore, one has
\vskip.1in
\noindent {\bf Proposition:} The matrix form of generators in the adjoint
representation of $U_q(A_2^{(1)})$ is given by
\begin{eqnarray}
&&h_\alpha={\rm diag}(1,-1,2,0,0,-2,1,-1)\,,~~~~h_\beta={\rm diag}(
  1,2,-1,0,0,1,-2,-1)\nonumber\\
&&E_\alpha=e_{12}+[2]_q^{1/2}e_{34}+[2]_q^{1/2}e_{46}+e_{78}\,,~~~~
  F_\alpha=e_{21}+[2]_q^{1/2}e_{43}+[2]_q^{1/2}e_{66}+e_{87}\nonumber\\
&&E_\beta=e_{13}+[2]_q^{-1/2}e_{24}+\left (\frac{[3]_q}{[2]_q}\right )^{1/2}
  e_{25}+[2]_q^{-1/2}e_{47}+\left (\frac{[3]_q}{[2]_q}\right )^{1/2}
  e_{57}+e_{68}\nonumber\\
&&F_\beta=e_{31}+[2]_q^{-1/2}e_{42}+\left (\frac{[3]_q}{[2]_q}\right )^{1/2}
  e_{52}+[2]_q^{-1/2}e_{74}+\left (\frac{[3]_q}{[2]_q}\right )^{1/2}
  e_{75}+e_{86}\nonumber\\
&&E_{\alpha+\beta}=-[2]_q^{-1/2}q^{-2}e_{14}+\left (\frac{[3]_q}{[2]_q}
  \right )^{1/2}e_{15}-q^{-1}e_{26}+e_{37}+[2]_q^{-1/2}qe_{48}
  -\left (\frac{[3]_q}{[2]_q}\right )^{1/2}q^{-1}e_{58}\nonumber\\
&&F_{\alpha+\beta}=-[2]_q^{-1/2}q^{2}e_{41}+\left (\frac{[3]_q}{[2]_q}
  \right )^{1/2}e_{51}-qe_{62}+e_{73}+[2]_q^{-1/2}q^{-1}e_{84}
  -\left (\frac{[3]_q}{[2]_q}\right )^{1/2}qe_{85}\nonumber\\
&&E'_{\beta+n\delta}=q^ne_{13}+[2]_q^{-1/2}q^ne_{24}+
  \left (\frac{[3]_q}{[2]_q}\right )^{1/2}q^{-3n}e_{25}+\nonumber\\
&&~~~~~~~~~~~+[2]_q^{-1/2}q^{-3n}e_{47}+\left (\frac{[3]_q}{[2]_q}\right
)^{1/2}
  q^ne_{57}+q^{-3n}e_{68}\nonumber\\
&&F'_{\beta+n\delta}=q^{-n}e_{31}+[2]_q^{-1/2}q^{-n}e_{42}+
  \left (\frac{[3]_q}{[2]_q}\right )^{1/2}q^{3n}e_{52}+\nonumber\\
&&~~~~~~~~~~~+[2]_q^{-1/2}q^{3n}e_{74}+\left (\frac{[3]_q}{[2]_q}\right )^{1/2}
  q^{-n}e_{75}+q^{3n}e_{86}\nonumber\\
&&E'_{(\delta-\beta)+n\delta}=-q^{n+2}e_{31}-[2]_q^{-1/2}q^{n+3}e_{42}-
  \left (\frac{[3]_q}{[2]_q}\right )^{1/2}q^{-3n-1}e_{52}\nonumber\\
&&~~~~~~~~~~~~~~~-[2]_q^{-1/2}q^{-3n-3}e_{74}-\left (
  \frac{[3]_q}{[2]_q}\right )^{1/2}
  q^{n+1}e_{75}-q^{-3n-2}e_{86}\nonumber\\
&&F'_{(\delta-\beta)+n\delta}=-q^{-n-2}e_{13}-[2]_q^{-1/2}q^{-n-3}e_{24}-
  \left (\frac{[3]_q}{[2]_q}\right )^{1/2}q^{3n+1}e_{25}\nonumber\\
&&~~~~~~~~~~~~~~~~-[2]_q^{-1/2}q^{3n+3}e_{47}-\left (
  \frac{[3]_q}{[2]_q}\right )^{1/2}
  q^{-n-1}e_{57}-q^{3n+2}e_{68}\nonumber\\
&&E^{'(\alpha)}_{n\delta}=[2]_q^{-1}(-1)^{n-1}\frac{[n]_q}{n}\left \{
  q^{-n}e_{11}-q^{-3n}e_{22}+(q^n+q^{-n})e_{33}+q^{-n}(q^{2n}-q^{-2n})e_{44}
  \right .\nonumber\\
&&~~~~~~~~~~~  \left .-q^{-2n}(q^n+q^{-n})e_{66}
  q^ne_{77}-q^{-n}e_{88}\right \}\nonumber\\
&&F^{'(\alpha)}_{n\delta}=[2]_q^{-1}(-1)^{n-1}\frac{[n]_q}{n}\left \{
  q^{n}e_{11}-q^{3n}e_{22}+(q^n+q^{-n})e_{33}-q^{n}(q^{2n}-q^{-2n})e_{44}
  \right .\nonumber\\
&&~~~~~~~~~~~~ \left . -q^{2n}(q^n+q^{-n})e_{66}
  q^{-n}e_{77}-q^{n}e_{88}\right \}\nonumber\\
&&E^{'(\beta)}_{n\delta}=-[2]_q^{-1}\frac{[n]_q}{n}\left \{
  q^{2n}e_{11}+(q^{2n}+q^{-2n})e_{22}-e_{33}-q^{-n}(q^n-q^{-n})e_{44}+
  \right .\nonumber\\
&&~~~~~ \left .+ q^{-n}(q^n-q^{-n})(q^{2n}
  +1+q^{-2n})e_{55}+q^{-2n}e_{66}-q^{-2n}(q^{2n}+q^{-2n})e_{77}-q^{-4n}e_{88}
  \right \}\nonumber\\
&&F^{'(\beta)}_{n\delta}=-[2]_q^{-1}\frac{[n]_q}{n}\left \{
  q^{-2n}e_{11}+(q^{2n}+q^{-2n})e_{22}-e_{33}+q^{n}(q^n-q^{-n})e_{44}
  \right .\nonumber\\
&&~~~~~ \left . -q^{n}(q^n-q^{-n})(q^{2n}
  +1+q^{-2n})e_{55}+q^{2n}e_{66}-q^{2n}(q^{2n}+q^{-2n})e_{77}-q^{4n}e_{88}
  \right \}\label{action}
\end{eqnarray}
\vskip.1in
\noindent {\bf Proof:} Straightforward calculations + induction in $n$.
\vskip.1in
\noindent {\bf Proposition:} We have the following properties for the
generators in (\ref{action}),
\begin{eqnarray}
&&(E_\alpha)^2=[2]_qe_{36}\,,~~~(E_\alpha)^3=0\,,~~~(F_\alpha)^2=[2]_qe_{63}
  \,,~~~(F_\alpha)^3=0\nonumber\\
&&(E_{\alpha+\beta})^2=-[2]_qq^{-1}e_{18}\,,~~(E_{\alpha+\beta})^3=0
  \,,~~(F_{\alpha+\beta})^2=-[2]_qqe_{63}
  \,,~~(F_{\alpha+\beta})^3=0\nonumber\\
&&(E'_{\beta+n\delta})^2=[2]_qq^{-2n}e_{27}\,,~~(E'_{\beta+n\delta})^3=0\,,~~
  (F'_{\beta+n\delta})^2=[2]_qq^{2n}e_{72}\,,~~
  (F'_{\beta+n\delta})^3=0\nonumber\\
&&(E'_{(\delta-\beta)+n\delta})^2=[2]_qq^{-2n}e_{72}\,,~~
  (E'_{(\delta-\beta)+n\delta})^3=0\nonumber\\
&&(F'_{(\delta-\beta)+n\delta})^2=[2]_qq^{2n}e_{27}\,,~~
  (F'_{(\delta-\beta)+n\delta})^3=0\label{nilpotent}
\end{eqnarray}
\vskip.1in
\noindent {\bf Proof:} Easily checked.
\vskip.1in
We now apply (\ref{loop-R}) to $V_{(8)}\otimes V_{(8)}$, where
$V_{(8)}$ is the adjoint representation of $U_q(A_2)$.
Inserting (\ref{action}) into (\ref{loop-R}), we see that in the expansion
of each $q$-exponential only three terms survive thanks to the celebrated
properties of generators, eq.(\ref{nilpotent}). Thus one is able to work
out the infinite products in (\ref{loop-R}).
The contributions from the imaginary root vectors in
(\ref{loop-R}) can also be worked out term by term and written as a very
compact form. The final result may be put in the explicit and
compact form,
\begin{eqnarray}
R_{(8),(8)}(x,y)&=&\left \{1+(q-q^{-1})\sum_{n=0}^\infty\left (\frac{x}{y}
  \right )^n \left (E'_{\alpha+n\delta}\otimes F'_{\alpha+n\delta}\right )
  +\right .\nonumber\\
& & \left . +[2]_qq^{-1}
  (q-q^{-1})^2\frac{y^2(y+q^4x)}{(y^2-x^2)(y-q^2x)}e_{36}\otimes
  e_{63}\right \}\nonumber\\
& &\cdot\left \{1+(q-q^{-1})\sum_{n=0}^\infty\left (\frac{x}{y}\right )^n
  \left (E'_{\alpha+\beta+n\delta}\otimes F'_{\alpha+\beta+n\delta}\right )
  +\right .\nonumber\\
& & \left . +[2]_qq^{-1}
  (q-q^{-1})^2\frac{y^2(y+q^4x)}{(y^2-x^2)(y-q^2x)}e_{18}\otimes
  e_{81}\right \}\nonumber\\
& &\cdot\left \{1+(q-q^{-1})\sum_{n=0}^\infty\left (\frac{x}{y}\right )^n
  \left (E'_{\beta+n\delta}\otimes F'_{\beta+n\delta}\right )
  +\right .\nonumber\\
& & \left . +[2]_qq^{-1}
  (q-q^{-1})^2\frac{y^2(y+q^4x)}{(y^2-x^2)(y-q^2x)}e_{27}\otimes
  e_{72}\right \}\nonumber\\
& &\cdot\left \{{\rm imaginary~root~vectors~contribution}\right \}\nonumber\\
& &\cdot\left \{1+(q-q^{-1})\sum_{n=0}^\infty\left (\frac{x}{y}\right )^{n+1}
  \left (E'_{(\delta-\beta)+n\delta}\otimes F'_{(\delta-\beta)+n\delta}\right )
  +\right .\nonumber\\
& & \left . +[2]_qq^{-1}
  (q-q^{-1})^2\frac{x^2(y+q^4x)}{(y^2-x^2)(y-q^2x)}e_{72}\otimes
  e_{27}\right \}\nonumber\\
& &\cdot\left \{1+(q-q^{-1})\sum_{n=0}^\infty\left (\frac{x}{y}\right )^{n+1}
  \left (E'_{(\delta-\alpha)+n\delta}\otimes F'_{(\delta-\alpha)+n\delta}
  \right )+\right .\nonumber\\
& & \left . +[2]_qq^{-1}
  (q-q^{-1})^2\frac{x^2(y+q^4x)}{(y^2-x^2)(y-q^2x)}e_{63}\otimes
  e_{36}\right \}\nonumber\\
& &\cdot\left \{1+(q-q^{-1})\sum_{n=0}^\infty\left (\frac{x}{y}\right )^{n+1}
  \left (E'_{(\delta-\alpha-\beta)+n\delta}\otimes
  F'_{(\delta-\alpha-\beta)+n\delta}\right )\right .+\nonumber\\
& &\left .+[2]_qq^{-1}(q-q^{-1})^2\frac{x^2(y+q^4x)}{(y^2-x^2)(y-q^2x)}
  e_{81}\otimes e_{18}\right \}\nonumber\\
& &\cdot\left \{q^2\sum_{i=1}^8(1-\delta_{i4}-\delta_{i5})(e_{ii}\otimes
e_{ii})
  +q(e_{11}\otimes e_{22}+e_{11}\otimes e_{33}+e_{22}\otimes e_{66}+
  \right .\nonumber\\
& &+ e_{33}\otimes e_{77}+e_{66}\otimes e_{88}+e_{77}\otimes e_{88}+
  \{\leftrightarrow\})
  +q^{-1}(e_{11}\otimes e_{66}+e_{11}\otimes e_{77}+\nonumber\\
& &+  e_{22}\otimes e_{33}+
  e_{22}\otimes e_{88}+e_{33}\otimes e_{88}+e_{66}\otimes e_{77}+\nonumber\\
& &  \{\leftrightarrow\})
  \left .+  q^{-2}(e_{11}\otimes e_{88}+e_{22}\otimes e_{77}+
  e_{33}\otimes e_{66}+\{\leftrightarrow\})\right \}\label{88R}
\end{eqnarray}
where $"\{\leftrightarrow\}"$ denotes the interchange of the quantities in
the space $X\otimes Y$\,;\,$E'_{\beta+n\delta}$\,,~$F'_{\beta+n\delta}$
\,,~$E'_{(\delta-\beta)+n\delta}$
\,,~$F'_{(\delta-\beta)+n\delta}$ are given in (\ref{action}) and
\begin{eqnarray}
&&E'_{\alpha+n\delta}=(-1)^n\left \{q^{-2n}e_{12}+[2]_q^{1/2}q^{-2n}e_{34}+
  [2]_q^{1/2}e_{46}+e_{78}\right \}\nonumber\\
&&F'_{\alpha+n\delta}=(-1)^n\left \{q^{2n}e_{21}+[2]_q^{1/2}q^{2n}e_{43}+
  [2]_q^{1/2}e_{64}+e_{87}\right \}\nonumber\\
&&E'_{\alpha+\beta+n\delta}=(-1)^n\left \{-[2]_q^{-1/2}q^{-2n-2}e_{14}+
  \left (\frac{[3]_q}{[2]_q}\right )^{1/2}q^{-2n}e_{15}-q^{-1}e_{26}
  +\right .\nonumber\\
&&~~~~~~~~~~~~~~~+\left .q^{-2n}e_{37}+[2]_q^{-1/2}qe_{48}-\left (
  \frac{[3]_q}{[2]_q}
  \right )^{1/2}q^{-1}e_{58}\right \}\nonumber\\
&&F'_{\alpha+\beta+n\delta}=(-1)^n\left \{-[2]_q^{-1/2}q^{2n+2}e_{41}+
  \left (\frac{[3]_q}{[2]_q}\right )^{1/2}q^{2n}e_{51}-qe_{62}
  +\right .\nonumber\\
&&~~~~~~~~~~~~~~~+\left .q^{2n}e_{73}+[2]_q^{-1/2}q^{-1}e_{84}-\left (
  \frac{[3]_q}{[2]_q}
  \right )^{1/2}qe_{85}\right \}\nonumber\\
&&E'_{(\delta-\alpha)+n\delta}=(-1)^n\left \{q^{-2n-1}e_{21}+
  [2]_q^{1/2}q^{-2n}e_{43}+
  [2]_q^{1/2}e_{64}+qe_{87}\right \}\nonumber\\
&&F'_{(\delta-\alpha)+n\delta}=(-1)^n\left \{q^{2n+1}e_{12}+
  [2]_q^{1/2}q^{2n}e_{34}+
  [2]_q^{1/2}e_{46}+q^{-1}e_{78}\right \}\nonumber\\
&&E'_{(\delta-\alpha-\beta)+n\delta}=(-1)^n\left \{-[2]_q^{-1/2}q^{-2n+2}e_{41}
  +\left (\frac{[3]_q}{[2]_q}\right )^{1/2}q^{-2n}e_{51}-q^2e_{62}
  +\right .\nonumber\\
&&~~~~~~~~~~~~~~~+\left .q^{-2n-1}e_{73}+[2]_q^{-1/2}q^{-1}e_{84}-\left (
  \frac{[3]_q}{[2]_q}
  \right )^{1/2}qe_{85}\right \}\nonumber\\
&&F'_{(\delta-\alpha-\beta)+n\delta}=(-1)^n\left \{-[2]_q^{-1/2}q^{2n-2}e_{14}
  +\left (\frac{[3]_q}{[2]_q}\right )^{1/2}q^{2n}e_{15}-q^{-2}e_{26}
  +\right .\nonumber\\
&&~~~~~~~~~~~~~~~+\left .q^{2n+1}e_{37}+[2]_q^{-1/2}qe_{48}-\left (
  \frac{[3]_q}{[2]_q}
  \right )^{1/2}q^{-1}e_{58}\right \}\nonumber\\
&&\{{\rm imaginary~root~vectors~contribution}\}=a'/a\,\sum_{i=1}^8\left (
  1+(b/b'-1)\delta_{i4}+(c/c'-1)\delta_{i5}\right )(e_{ii}\otimes e_{ii})+
  \nonumber\\
&&~~~~~~~~~~~~~~~~+  a'(e_{11}\otimes e_{22}+e_{11}\otimes e_{33})
  +aa'(e_{11}\otimes e_{44})
  +a'c(e_{11}\otimes e_{55})+a(e_{11}\otimes e_{66})+\nonumber\\
&&~~~~~~~~~~~~~~~~+aa'c(e_{11}\otimes
  e_{77})+ac(e_{11}\otimes e_{88})+1/a(e_{22}\otimes e_{11})+
  1/b'(e_{22}\otimes e_{33})+\nonumber\\
&&~~~~~~~~~~~~~~~~+  a'/b'(e_{22}\otimes e_{44})+a'c(e_{22}\otimes
  e_{55})+a'(e_{22}\otimes e_{66})+aa'c/b'(e_{22}\otimes e_{77})+\nonumber\\
&&~~~~~~~~~~~~~~~~+aa'c(e_{22}\otimes e_{88})+1/a(e_{33}\otimes e_{11})
  +b(e_{33}\otimes e_{22})
  +a'b(e_{33}\otimes e_{44})+\nonumber\\
&&~~~~~~~~~~~~~~~~+  ab(e_{33}\otimes e_{66})+a'(e_{33}\otimes e_{77})+
  a(e_{33}\otimes e_{88})+1/(aa')(e_{44}\otimes e_{11})+\nonumber\\
&&~~~~~~~~~~~~~~~~+  b/a(e_{44}\otimes
  e_{22})+1/(ab')(e_{44}\otimes e_{33})+a'b(e_{44}\otimes e_{66})+a'/b'
  (e_{44}\otimes e_{77})+\nonumber\\
&&~~~~~~~~~~~~~~~~+aa'(e_{44}\otimes e_{88})+1/(ac')(e_{55}\otimes e_{11}+
  e_{55}\otimes e_{22})+a'c(e_{55}\otimes e_{77}+e_{55}\otimes e_{88})
  +\nonumber\\
&&~~~~~~~~~~~~~~~~+  1/a'(e_{66}\otimes e_{11})+
  1/a(e_{66}\otimes e_{22})+1/(a'b')(e_{66}\otimes e_{33})+1/(ab')(e_{66}
  \otimes e_{44})+\nonumber\\
&&~~~~~~~~~~~~~~~~+  1/b'(e_{66}\otimes e_{77})+a'(e_{66}\otimes e_{88})+
  1/(aa'c')(e_{77}\otimes e_{11})+
  b/(aa'c')(e_{77}\otimes e_{22})+\nonumber\\
&&~~~~~~~~~~~~~~~~+  1/a(e_{77}\otimes e_{33})+b/a(e_{77}
  \otimes e_{44})+1/(ac')(e_{77}\otimes e_{55})+b(e_{77}\otimes e_{66})+
  \nonumber\\
&&~~~~~~~~~~~~~~~~+  a'(e_{77}\otimes e_{88})+
  1/(a'c')(e_{88}\otimes e_{11})+1/(aa'c')(e_{88}\otimes e_{22})+
  1/a'(e_{88}\otimes e_{33})+\nonumber\\
&&~~~~~~~~~~~~~~~~+  1/(aa')(e_{88}\otimes e_{44})+1/(ac')(e_{88}\otimes
  e_{55})+1/a(e_{88}\otimes e_{66}+e_{88}\otimes e_{77})\label{generators}
\end{eqnarray}
in which we have defined
\begin{eqnarray}
&&a=\frac{y-q^2x}{y-x}\,,~~~a'=\frac{y-q^{-2}x}{y-x}\,,~~~b=\frac{y-q^4x}
  {y-q^2x}\nonumber\\
&&b'=\frac{y-q^{-4}x}{y-q^{-2}x}\,,~~~c=\frac{y-q^6x}{y-q^4x}\,,~~~
  c'=\frac{y-q^{-6}x}{y-q^{-4}x}
\end{eqnarray}
We see that (\ref{88R}) is an extreamly explicit formula: the sums in
(\ref{88R}) can be easily worked out. We do this in the Appendix B.

\section{Concluding Remarks}
In this paper we have given a detailed account for the results presented
in our previous short letter \cite{ZG2} where only the results have
been announced.

We believe that along our line
we may at least search for the solution to the following problems.
Firstly, we may try to extend the above to other types of
quantum affine algebras (twisted or nontwisted).
To this effect, we first have to answer the
question how to quantize loop representations of the other types
(type B, C, D, E and extotic) of groups. Secondly, we may
wonder if there exist some kind of "universal" integrable lattice
models which have our spectral-dependent $R$-matrix as their Boltzmann
weights. Thirdly, we may consider the possibility of finding and computing
eigenvalues of Casimir operators constructed from these sepctral
parameter dependent $R$-matrix which are expected to play some role in
one dimensional open spin chains \cite{Cuerno}\cite{Davies}.
Finally, we believe
our formula will be useful in quantizing the conformal affine Toda
theories \cite{Bonora} and in the recently-developed $q$-deformed
WZNW CFT's \cite{Frenkel}\cite{Japanese}.
These are problems now under consideration.

\vskip.3in
\begin{center}
{\bf Acknowledgements:}
\end{center}
Y.Z.Z. would like to thank Anthony John Bracken for contineous interest,
suggestions and discussions, to thank V.N.Tolstoy for communication of
his papers on quantum groups and to thank R.Cuerno for email correspondences.
The financial support from Australian Research Council
is gratefully acknowledged.

\section{Appendix A}
We consider here finite-dimensional loop representations of
$U_q(gl(n)^{(1)})=U_q(A_{n-1}^{(1)})$
with the Chevalley generators $\{E_i,F_i,q^{h_i},0\leq i<n;q^d\}$ in which
\begin{equation}
E_i\equiv E_{i\,i+1}\,,~~~F_i\equiv E_{i+1\,i}\,,~~~q^{h_i}\,,~~~\,h_i
\equiv E_{i\,i}-E_{i+1\,i+1}\,,\,1\leq i< n\,,~~~\,q^{E_{nn}}
\end{equation}
are the usual Chevalley generators of $U_q(gl(n))$.
We define
\begin{eqnarray}
&&E_{ij}=E_{ik}E_{kj}-q^{-1}E_{kj}E_{ik}\,,~~~~i<k<j\nonumber\\
&&E_{ij}=E_{ik}E_{kj}-qE_{kj}E_{ik}\,,~~~~i>k>j
\end{eqnarray}
and put
\begin{equation}
E_\psi\equiv q^{E_{11}+E_{nn}}E_{1n}\,,~~~~
F_\psi\equiv E_{n1}q^{-E_{11}-E_{nn}}\,,~~~~
h_\psi\equiv E_{11}-E_{nn}\label{evaluation}
\end{equation}
then we have
\vskip.1in
\noindent {\bf Proposition:} For any given $z\in {\bf C}^\times$, there is
a homomorphism of algebras ${\rm ev}_z$: $U_q(gl(n)^{(1)})\rightarrow
U_q(gl(n))$, in terms of the Chevalley generators,
\begin{eqnarray}
&&{\rm ev}_z(E_i)=E_i\,,~~~{\rm ev}_z(F_i)=F_i\,,~~~{\rm ev}_z(h_i)=h_i
  \nonumber\\
&&{\rm ev}_z(E_0)=zF_\psi\,,~~~{\rm ev}_z(F_0)=z^{-1}E_\psi\,,~~~
  {\rm ev}_z(h_0)=-h_\psi\,,~~~{\rm ev}_z(c)=0
\end{eqnarray}
\vskip.1in
\noindent {\bf Proof:} To show they define a homomorphism $U_q(gl(n)^{(1)})
\rightarrow U_q(gl(n))$, one needs to check that the relations in
(\ref{relations1}) are satisfied. This is immediate except for the last
two, which reduce to
\begin{eqnarray}
&&({\rm ad}_{q^{-1}}F_i)^{1+(\psi,\alpha_i)}F_0= ({\rm ad}_{q^{-1}}
  F_i)^{1+(\psi,\alpha_i)}E_\psi=0\label{1}\\
&&({\rm ad}_{q^{-1}}F_0)^{1+(\psi,\alpha_i)}F_i= ({\rm ad}_{q^{-1}}
  E_\psi)^{1+(\psi,\alpha_i)}F_i=0\label{1'}
\end{eqnarray}
and two similar relations with interchange $F_i\leftrightarrow E_i\,,\,
F_0\leftrightarrow E_0\,,\,q^{-1}\leftrightarrow q$. We now prove (\ref{1}).
First we consider the case: $1<i<i+1<n$. In this case $(\psi,\alpha_i)=0$,
and the l.h.s. of (\ref{1}) becomes
\begin{equation}
({\rm ad}_{q^{-1}}F_i)E_\psi=[F_i,E_\psi]=q^{E_{11}+E_{nn}}[E_{i+1\,i},
E_{1n}]
\end{equation}
which can be easily checked to be vanishing. We then consider the $i=1$ case.
In this case the l.h.s. of (\ref{1}) reads
\begin{equation}
({\rm ad}_{q^{-1}}F_1)^2E_\psi=({\rm ad}_{q^{-1}}E_{21})({\rm
ad}_{q^{-1}}E_{21}
)E_\psi\label{2}
\end{equation}
One can easily show $({\rm ad}_{q^{-1}}E_{21})E_\psi=q^{E_{22}+E_{nn}}E_{2n}$.
Inserting this into (\ref{2}), one get
\begin{equation}
(\ref{2})=({\rm ad}_{q^{-1}}E_{21})q^{E_{22}+E_{nn}}E_{2n}=
q^{E_{22}+E_{nn}-1}[E_{21}\,,\,E_{2n}]=0
\end{equation}
as required. Finally for
$i=n$, we see the l.h.s. of (\ref{1}) reduces to
\begin{equation}
({\rm ad}_{q^{-1}}F_n)^2E_\psi=({\rm ad}_{q^{-1}}E_{n\,n-1})
({\rm ad}_{q^{-1}}E_{n\,n-1})E_\psi\label{3}
\end{equation}
Some direct computations give
\begin{equation}
(\ref{3})=q^{E_{11}+E_{nn}-1}\left \{q^{-1}E_{n\,n-1}[E_{n\,n-1},E_{1n}]
-q[E_{n\,n-1},E_{1n}E_{n\,n-1}\right \}
\end{equation}
which, using the directly checkable formula,
\begin{equation}
[E_{n\,n-1},E_{1n}]=-q^{E_{n-1\,n-1}-E_{nn}}E_{1\,n-1}\,,
\end{equation}
is easily seen to be vanishing. We may similarly prove (\ref{1'}).
\vskip.1in
\noindent {\bf Remark:} Since $N\equiv \sum_{i=1}^n\,E_{ii}$ commutes with
everything, therefore, if we set, instead of (\ref{evaluation}),
\begin{equation}
E_\psi\equiv q^{E_{11}+E_{nn}-\frac{2}{n}N}\,E_{1n}\,,~~~~
F_\psi\equiv E_{n1}\,q^{-E_{11}-E_{nn}+\frac{2}{n}N}\,E_{1n}\label{4}
\end{equation}
then the above proposition in this appendix still holds. It turns out
that it is more convenient to use (\ref{4}) as we did in the previous
sections.

\section{Appendix B}
For completeness, in this appendix we work out the sums appearing in
(\ref{88R}). We list the results below:
\begin{eqnarray}
&&{\rm the~first~sum}=\frac{y}{y-x}\{e_{12}\otimes e_{21}
  +[2]_q^{1/2}(e_{12}\otimes e_{43})+[2]_q^{1/2}(e_{34}\otimes e_{21})
  +[2]_q(e_{34}\otimes e_{43})+\nonumber\\
&&~~~~~~~~~+[2]_q(e_{46}\otimes e_{64})+[2]_q^{1/2}(e_{46}\otimes
  e_{87})+[2]_q^{1/2}(e_{78}\otimes e_{64})+e_{78}\otimes e_{87}\}+\nonumber\\
&&~~~~~~~~~+\frac{y}{y-q^{-2}x}\{[2]_q^{1/2}(e_{12}\otimes
  e_{64})+e_{12}\otimes e_{87}+[2]_q(e_{34}\otimes e_{64})+[2]_q^{1/2}
  (e_{34}\otimes e_{87})\}+\nonumber\\
&&~~~~~~~~~+\frac{y}{y-q^2x}\{[2]_q^{1/2}(e_{46}\otimes
  e_{21})+[2]_q(e_{46}\otimes e_{43})+e_{78}\otimes e_{21}+[2]_q^{1/2}
  (e_{78}\otimes e_{43})\}\nonumber\\
&&{\rm the~second~sum}=\frac{y}{y-x}\{1/[2]_q(e_{14}\otimes e_{41})-[3]_q^{1/2}
  /(q^2[2]_q)(e_{14}\otimes e_{51})-1/(q^2[2]_q^{1/2})(e_{14}\otimes e_{73})
  \nonumber\\
&&~~~~~~~~~ -q^2[3]_q^{1/2}/[2]_q(e_{15}\otimes e_{41})
  +[3]_q/[2]_q(e_{15}\otimes e_{51})+([3]_q/[2]_q)^{1/2}
  (e_{15}\otimes e_{73})+\nonumber\\
&&~~~~~~~~~+  e_{26}\otimes e_{62}
  -1/(q^2[2]_q^{1/2})(e_{26}\otimes e_{84})
  +([3]_q/[2]_q)^{1/2}(e_{26}\otimes e_{85})\nonumber\\
&&~~~~~~~~~  -q^2/[2]_q^{1/2}
  (e_{37}\otimes e_{41})+([3]_q/[2]_q)^{1/2}(e_{37}\otimes e_{51})+\nonumber\\
&&~~~~~~~~~  +e_{37}\otimes e_{73}
  -q^2/[2]_q^{1/2}(e_{48}\otimes e_{62})+e_{48}\otimes
  e_{84}-q^2[3]_q^{1/2}/[2]_q(e_{48}\otimes e_{85})\nonumber\\
&&~~~~~~~~~  +([3]_q/[2]_q)^{1/2}
  (e_{58}\otimes e_{62})
  -[3]^{1/2}/(q^2[2]_q)(e_{58}\otimes e_{84})
  +[3]_q/[2]_q(e_{58}\otimes e_{85})\}\nonumber\\
&&~~~~~~~~~  +\frac{y}{y-q^{-2}x}\{1/(q[2]_q^{1/2})
  (e_{14}\otimes e_{62})-1/(q^3[2]_q)
  (e_{14}\otimes e_{84})
  +[3]_q^{1/2}/(q[2]_q)(e_{14}\otimes e_{85})\nonumber\\
&&~~~~~~~~~  -q^{-1}([3]_q/[2]_q
  )^{1/2}(e_{15}\otimes e_{62})+[3]_q^{1/2}/(q[2]_q)(e_{15}\otimes e_{84})-
  q([3]_q/[2]_q)(e_{15}\otimes e_{85})\nonumber\\
&&~~~~~~~~~-q(e_{37}\otimes e_{62})
  +1/(q[2]_q^{1/2})(e_{37}\otimes e_{84})
  -q([3]_q/[2]_q)^{1/2}(e_{37}\otimes e_{85})\}\nonumber\\
&&~~~~~~~~  +\frac{y}{y-q^2x}\{q/[2]_q^{1/2}
  (e_{26}\otimes e_{41})
  -q^{-1}([3]_q/[2]_q)^{1/2}(e_{26}\otimes e_{51})-q^{-1}
  (e_{26}\otimes e_{73})\nonumber\\
&&~~~~~~~~  -q^3/[2]_q(e_{48}\otimes e_{41})+q[3]_q^{1/2}/[2]_q
  (e_{48}\otimes e_{51})
  +q/[2]_q^{1/2}(e_{48}\otimes e_{73})\nonumber\\
&&~~~~~~~~  +q[3]_q^{1/2}/[2]_q(e_{58}
  \otimes e_{41})-q^{-1}([3]_q/[2]_q)(e_{58}\otimes e_{51})-q^{-1}([3]_q/[2]_q)
  ^{1/2}(e_{58}\otimes e_{73})\}\nonumber\\
&&{\rm the~third~sum}=\frac{y}{y-x}\{e_{13}\otimes e_{31}+1/[2]_q^{1/2}(e_{13}
  \otimes e_{42})+([3]_q/[2]_q)^{1/2}(e_{13}\otimes e_{75})+\nonumber\\
&&~~~~~~~~+  1/[2]_q^{1/2}(e_{24}\otimes e_{31})
  +1/[2]_q(e_{24}\otimes e_{42})+[3]_q^{1/2}/[2]_q(e_{24}
  \otimes e_{75})+\nonumber\\
&&~~~~~~~~+  [3]_q/[2]_q(e_{25}\otimes e_{52})+[3]_q^{1/2}/[2]_q
  (e_{25}\otimes e_{74})
  +([3]_q/[2]_q)^{1/2}(e_{25}\otimes e_{86})+\nonumber\\
&&~~~~~~~~+[3]_q^{1/2}/[2]_q(e_{47}\otimes e_{52})+
  1/[2]_q(e_{47}\otimes e_{74})+\nonumber\\
&&~~~~~~~~+  1/[2]_q^{1/2}
  (e_{47}\otimes e_{86})+([3]_q/[2]_q)^{1/2}(e_{57}\otimes e_{31})+
  [3]_q^{1/2}/[2]_q(e_{57}\otimes e_{42})+\nonumber\\
&&~~~~~~~~+[3]_q/[2]_q(e_{57}\otimes e_{75})+([3]_q/[2]_q)^{1/2}
  (e_{68}\otimes e_{52})+1/[2]_q^{1/2}(e_{68}\otimes e_{74})+e_{68}\otimes
  e_{86}\}\nonumber\\
&&~~~~~~~~+\frac{y}{y-q^4x}\{([3]_q/[2]_q)^{1/2}(e_{13}\otimes e_{52})
  +1/[2]_q^{1/2}(e_{13}\otimes e_{74})+e_{13}\otimes e_{86}+\nonumber\\
&&~~~~~~~~+  [3]_q^{1/2}/[2]_q
  (e_{24}\otimes e_{52})
  +1/[2]_q(e_{24}\otimes e_{74})+1/[2]_q^{1/2}(e_{24}\otimes
  e_{86})+\nonumber\\
&&~~~~~~~~+[3]_q/[2]_q(e_{57}\otimes e_{52})+[3]_q^{1/2}/[2]_q(e_{57}\otimes
  e_{74})+([3]_q/[2]_q)^{1/2}(e_{57}\otimes e_{86})\}+\nonumber\\
&&~~~~~~~~+  \frac{y}{y-q^{-4}x}\{([3]_q/[2]_q)^{1/2}(e_{25}\otimes e_{31})
  +[3]_q^{1/2}/[2]_q(e_{25}\otimes e_{42})
  +[3]_q/[2]_q(e_{25}\otimes e_{75})\nonumber\\
&&~~~~~~~~  +1/[2]_q^{1/2}(e_{47}\otimes e_{31})+1/[2]_q(e_{47}\otimes
  e_{42})+[3]_q^{1/2}/[2]_q(e_{47}\otimes e_{75})+\nonumber\\
&&~~~~~~~~  +e_{68}\otimes e_{31}
  +1/[2]_q^{1/2}(e_{68}\otimes e_{42})+([3]_q/[2]_q)^{1/2}
  (e_{68}\otimes e_{75})\}\nonumber\\
&&{\rm the~fourth~sum}=\frac{x}{y-x}\{e_{31}\otimes e_{13}+1/(q[2]_q^{1/2})
  (e_{31}\otimes e_{24})+q([3]_q/[2]_q)^{1/2}(e_{31}\otimes e_{57})\nonumber\\
&&~~~~~~~~+  q/[2]_q^{1/2}(e_{42}\otimes e_{13})
  +1/[2]_q(e_{42}\otimes e_{24})+q^2[3]_q^{1/2}/[2]_q
  (e_{42}\otimes e_{57})+\nonumber\\
&&~~~~~~~~+  [3]_q/[2]_q(e_{52}\otimes e_{25})+q^2[3]_q^{1/2}/
  [2]_q(e_{52}\otimes e_{47})
  +q([3]_q/[2]_q)^{1/2}(e_{52}\otimes e_{68})+\nonumber\\
&&~~~~~~~~+  [3]_q^{1/2}
  /(q^2[2]_q)(e_{74}\otimes e_{25})+1/[2]_q(e_{74}\otimes e_{47})+
  1/(q[2]_q^{1/2})(e_{74}\otimes e_{68})+\nonumber\\
&&~~~~~~~~+q^{-1}([3]_q/[2]_q)^{1/2}(e_{75}\otimes e_{13})+
  [3]_q^{1/2}/(q^2[2]_q)(e_{75}\otimes e_{24})+\nonumber\\
&&~~~~~~~~  +[3]_q/[2]_q(e_{75}\otimes e_{57})+
  +q^{-1}([3]_q/[2]_q)^{1/2}(e_{86}\otimes e_{25})+
  q/[2]_q^{1/2}(e_{86}\otimes e_{47})+\nonumber\\
&&~~~~~~~~+  e_{86}\otimes e_{68}\}
  + \frac{x}{y-q^4x}\{q^3([3]_q/[2]_q)^{1/2}
  (e_{31}\otimes e_{25})
  +q^5/[2]_q^{1/2}(e_{31}\otimes e_{47})+\nonumber\\
&&~~~~~~~~+  q^4(e_{31}\otimes  e_{68})
  +q^4[3]_q^{1/2}/[2]_q(e_{42}\otimes e_{25})
  +q^6/[2]_q(e_{42}\otimes e_{47})+\nonumber\\
&&~~~~~~~~+q^5/[2]_q^{1/2}(e_{42}\otimes e_{68})+q^2[3]_q/[2]_q(e_{75}\otimes
  e_{25})+q^4[3]_q^{1/2}/
  [2]_q(e_{75}\otimes e_{47})\nonumber\\
&&~~~~~~~~+q^3([3]_q/[2]_q)^{1/2}(e_{75}\otimes e_{68})\}
  +\frac{x}{y-q^{-4}x}\{q^{-3}([3]_q/[2]_q)^{1/2}
  (e_{52}\otimes e_{13})\nonumber\\
&&~~~~~~~~+  [3]_q^{1/2}/(q^4[2]_q)(e_{52}\otimes e_{24})+
  +[3]_q/(q^2[2]_q)(e_{52}\otimes e_{57})+\nonumber\\
&&~~~~~~~~  +1/(q^5[2]_q^{1/2})
  (e_{74}\otimes e_{13})+1/(q^6[2]_q)(e_{74}\otimes e_{24})+[3]_q^{1/2}/
  (q^4[2]_q)(e_{74}\otimes e_{57})+\nonumber\\
&&~~~~~~~~+q^{-4}(e_{86}\otimes e_{13})+1/(q^5[2]_q^{1/2})(e_{86}
  \otimes e_{24})+q^{-3}([3]_q/[2]_q)^{1/2}(e_{86}\otimes e_{57})\nonumber\\
&&{\rm the~fifth~sum}=\frac{x}{y-x}\{e_{21}\otimes e_{12}+[2]_q^{1/2}/q
  (e_{21}\otimes e_{34})+q[2]_q^{1/2}(e_{43}\otimes e_{12})+[2]_q
  (e_{43}\otimes e_{34})\nonumber\\
&&~~~~~~~~+[2]_q(e_{64}\otimes e_{46})+[2]_q^{1/2}/q(e_{64}\otimes e_{78})+
  q[2]_q^{1/2}(e_{87}\otimes e_{46})+e_{87}\otimes e_{78}\}\nonumber\\
&&~~~~~~~+\frac{x}{y-q^{-2}x}\{[2]_q^{1/2}/q(e_{21}\otimes e_{46})+
  1/q^2(e_{21}\otimes e_{78})+[2]_q(e_{43}\otimes e_{46})+[2]_q^{1/2}/q
  (e_{43}\otimes e_{78})\}\nonumber\\
&&~~~~~~~~+\frac{x}{y-q^2x}\{q[2]_q^{1/2}(e_{64}\otimes e_{12})+
  [2]_q(e_{64}\otimes e_{34})+q^2(e_{87}\otimes e_{12})+q[2]_q^{1/2}
  (e_{87}\otimes e_{34})\}\nonumber\\
&&{\rm  the~sixth~sum}=\frac{x}{y-x}\{1/[2]_q(e_{41}\otimes e_{14})
  -q^2[3]_q^{1/2}/[2]_q(e_{41}\otimes e_{15})-q^3/[2]_q^{1/2}
  (e_{41}\otimes e_{37})\nonumber\\
&&~~~~~~~~-[3]_q^{1/2}/(q^2[2]_q)(e_{51}\otimes e_{14})+[3]_q/[2]_q
  (e_{51}\otimes e_{15})+q([3]_q/[2]_q)^{1/2}(e_{51}\otimes e_{37})
  \nonumber\\
&&~~~~~~~~+e_{62}\otimes e_{26}-q^3/[2]_q^{1/2}(e_{62}\otimes e_{48})+
  q([3]_q/[2]_q)^{1/2}(e_{62}\otimes e_{58})-1/(q^3[2]_q^{1/2})
  (e_{73}\otimes e_{14})\nonumber\\
&&~~~~~~~~+q^{-1}([3]_q/[2]_q)^{1/2}(e_{73}\otimes e_{15})+e_{73}\otimes
  e_{37}-1/(q^3[2]_q^{1/2})(e_{84}\otimes e_{26})\nonumber\\
&&~~~~~~~~+1/[2]_q(e_{84}\otimes e_{48})-[3]_q^{1/2}/(q^2[2]_q)(e_{84}
  \otimes e_{58})+q^{-1}([3]_q/[2]_q)^{1/2}(e_{85}\otimes e_{26})\nonumber\\
&&~~~~~~~~-q^2[3]_q^{1/2}/[2]_q(e_{85}\otimes e_{48})+[3]_q/[2]_q
  (e_{85}\otimes e_{58})\}+\frac{x}{y-q^{-2}x}\{1/[2]_q^{1/2}
  (e_{41}\otimes e_{26})\nonumber\\
&&~~~~~~~~-q^3/[2]_q(e_{41}\otimes e_{48})+q[3]_q^{1/2}/[2]_q(e_{41}
  \otimes e_{58})-q^{-2}([3]_q/[2]_q)^{1/2}(e_{51}\otimes e_{26})\nonumber\\
&&~~~~~~~~+q[3]_q^{1/2}/[2]_q(e_{51}\otimes e_{48})-q^{-1}([3]_q/[2]_q)
  (e_{51}\otimes e_{58})-1/q^3(e_{73}\otimes e_{26})\nonumber\\
&&~~~~~~~~+1/[2]_q^{1/2}(e_{73}\otimes e_{48})-q^{-2}([3]_q/[2]_q)^{1/2}
  (e_{73}\otimes e_{58})\}+\frac{x}{y-q^2x}\{1/[2]_q^{1/2}(e_{62}\otimes
  e_{14})\nonumber\\
&&~~~~~~~~-q^2([3]_q/[2]_q)^{1/2}(e_{62}\otimes e_{15})-q^3(e_{62}\otimes
  e_{37})-1/(q^3[2]_q)(e_{84}\otimes e_{14})\nonumber\\
&&~~~~~~~~+[3]_q^{1/2}/(q[2]_q)(e_{84}\otimes e_{15})+1/[2]_q^{1/2}(e_{84}
  \otimes e_{37})+[3]_q^{1/2}/(q[2]_q)(e_{85}\otimes e_{14})\nonumber\\
&&~~~~~~~~-q([3]_q/[2]_q)(e_{85}\otimes e_{15})-q^2([3]_q/[2]_q)^{1/2}
  (e_{85}\otimes e_{37})\}
\end{eqnarray}



\newpage
\begin{thebibliography}{99}
\bibitem{ZG2} Y.-Z.Zhang and M.D.Gould, {\em Quantum Affine Algebra and
  Universal $R$-Matrix with Spectral Parameter}, The University of
  Queensland preprint UQMATH-93-05, 1993, hep-th/9307007
\bibitem{Drinfeld} V.G.Drinfeld, {\em Proc. ICM, Berkeley} {\bf 1} (1986) 798
\bibitem{Jimbo} M.Jimbo, {\em Lett.Math.Phys.} {\bf 10} (1985) 63,
  {\em ibid} {\bf 11} (1986) 247;
  {\em Commun.Math.Phys.} {\bf 102} (1986) 247
\bibitem{Reshetikhin} N.Reshetikhin, {\em Quantized universal enveloping
  algebras, the Yang-Baxter equation and inveriants of links: I, II},
  preprints LOMI E-4-87, E-17-87; A.N.Kirillov and N.Reshetikhin,
  {\em Representations of the algebra $U_q(sl(2))$, $q$-orthogonal
  polynomials and invariants of links}, preprint LOMI E-9-88
\bibitem{Sierra} L.Alvarez-Gaum\'e, C.Gomes and G.Sierra, {\em Phys.Lett.}
   {\bf B220} 142; {\em Nucl.Phys.} {\bf B319} (1989) 155;
   G.Moore and N.Yu.Reshetikhin, {\em Nucl.Phys.} {\bf B328} (1989) 557
\bibitem{Witten} E.Witten, {\em Commun.Math.Phys.} {\bf 121} (1989) 351
\bibitem{ZGB2} R.B.Zhang, M.D.Gould and A.J.Bracken, {\em Commun.Math.Phys.}
  {\bf 137} (1991) 13; J.R.Links, M.D.Gould and R.B.Zhang, {\em
  Quantum supergroups, link polynomials and representation of the braid
  generator}, {\em Rev.Math.Phys.} (in press)
\bibitem{Faddeev} L.D.Faddeev, {\em Integrable models in (1+1)-dimensional
   quantum field theory}, in "{\em Recent advances in field theory and
   statistical mechanics}", p.563, North-Holland: Elsevier, 1984
\bibitem{Baxter} R.J.Baxter, {\em Exactly solved models in statistical
  mechanics}, Academic Press, New York, 1982
\bibitem{Wadati} M.Wadati, T.Deguchi and Y.Akutsu, {\em Phys.Rep.}
  {\bf 180} (1989) 247
\bibitem{Jones} V.F.R.Jones, {\em Int.J.Mod.Phys.} {\bf B4} (1990) 701
\bibitem{ZGB} R.B.Zhang, M.D.Gould and A.J.Bracken, {\em Nucl.Phys.}
  {\bf 354} (1991) 625
\bibitem{KT} S.M.Khoroshkin and U.N.Tolstoy, {\em Lett.Math.Phys.}
  {\bf 24} (1992) 231; {\em Funkz.Analyz. i ego Pril.} {\bf 26} (1992) 85;
  {\em The Cartan-Weyl basis and the universal $R$-matrix for quantum
  KM algebras and superalgebras}, in {\em Prof. Second Winger Symposium},
  Goslar, Germany (July 1991), to appear in {\em Lect. Notes in Phys.}
  (1992)
\bibitem{Bonora} O.Babelon and L.Bonora, {\em Phys.Lett.} {\bf 244}
  (1990) 220; F.Toppan and Y.-Z.Zhang, {\em Phys.Lett.} {\bf 292}
  (1992) 67
\bibitem{Kac} V.G.Kac, {\em Infinite dimensional Lie algebras},
   {\em Prog.Math.} {\bf 44}, Birkh\"auser, Boston, 1983
\bibitem{ZG} Y.-Z.Zhang and M.D.Gould, {\em On universal $R$-Matrix for
  quantized nontwisted rank 3 affine Lie algebras},
  The University of Queensland preprint, UQMATH-93-01, 1993, hep-th/9303095;
  {\em Unitarity and complete reducibility of certain modules over quantized
  affine Lie algebras}, The University of Queensland preprint, UQMATH-93-02,
  1993, hep-th/9303096, {\em J.Math.Phys.} (in press)
\bibitem{Cuerno} R.Cuerno, {\em private e-mail communications}
\bibitem{Davies} B.Davies,
  O.Foda, M.Jimbo, T.Miwa and A.Nakayashiki, {\em Commun.Math.Phys.}
  {\bf 151} (1993) 89; D.Altschuler and B.Davies, {\em Quantum loop
  modules and quantum spain chains}, hep-th/9305135
\bibitem{Frenkel} I.B.Frenkel and N.Reshetikhin, {\em Commun.Math.Phys.}
  {\bf 146} (1992) 1
\bibitem{Japanese} H.Awata, S.Odake and J.Shiraishi,
  {\em Free boson realization
  of $U_q(sl(n)^{(1)})$}, preprint RIMS-924, YITP/K-1018 and references
therein;
  A.Matsuo, {\em Free field realization of $q$-deformed primary fields for
  $U_q(\hat{sl2})$}, Nagoya preprint, 1992 and references therein;
  A.H.Bougourzi and R.A.Weston, {\em Matrix elements of $U_q(su(2)_k)$
  vertex operators via bosonization}, preprint CRM-1875
\end{thebibliography}
\end{document}